\RequirePackage{fix-cm}
\documentclass[twocolumn,epjc3]{svjour3}  
\smartqed  
\RequirePackage{graphicx}
\journalname{Eur. Phys. J. C}
\usepackage[dvipsnames]{xcolor}
\usepackage{amssymb}
\usepackage{cite}
\usepackage{amsmath}
\usepackage[normalem]{ulem}
\usepackage{soul}
\usepackage{graphicx}
\usepackage{dcolumn}
\usepackage{bm}
\usepackage{epstopdf}
\usepackage{mathrsfs}
\usepackage{amssymb,amsmath,amsfonts,latexsym}
\usepackage{nicefrac}
\usepackage[colorlinks=true,linktocpage=true,linkcolor=blue,citecolor=blue]{hyperref}
\usepackage[utf8]{inputenc}
\usepackage{lipsum}
\usepackage{slashed}
\usepackage{mathtools}
\usepackage{multirow}
\usepackage{caption}
\usepackage{subcaption}
\usepackage{appendix}

\def  \bea  {\begin{eqnarray}}
\def  \eea  {\end{eqnarray}}
\def  \nn   {\nonumber}
\def  \qperp   {q_{\perp}}
\def  \kperp   {k_{\perp}}
\def  \pperp   {p_{\perp}}
\def  \ppperp   {p'_{\perp}}

\def  \vecnabla   {\vec{\nabla}}
\def \vecB {\vec{B}}

\def  \l {{\left}}
\def \r {{\right}}

\def \l {\left}
\def \r {\right}

\newcommand{\stkout}[1]{\ifmmode\text{\sout{\ensuremath{#1}}}\else\sout{#1}\fi}

\begin{document}

\title{Dynamically screened  strongly quantized electron transport in binary neutron-star merger
}
\author{Sreemoyee Sarkar\thanksref{e1,addr1}
        \and
        Souvik Priyam Adhya \thanksref{e2,addr2}
}

\thankstext{e1}{sreemoyee.sarkar@nmims.edu}
\thankstext{e2}{souvik.adhya@ifj.edu.pl}

\institute{Mukesh Patel School of Technology Management and Engineering, NMIMS University, Mumbai-400056, India\label{addr1}
           \and
           Institute of Nuclear Physics, Polish Academy of Sciences,
  ul.\ Radzikowskiego 152, 31-342 Krakow, Poland \label{addr2}
}



\date{Received: date / Accepted: date}
\def  \bea  {\begin{eqnarray}}
\def  \eea  {\end{eqnarray}}
\def  \nn   {\nonumber}
\def  \qperp   {q_{\perp}}
\def  \kperp   {k_{\perp}}
\def  \pperp   {p_{\perp}}
\def  \ppperp   {p'_{\perp}}

\def  \vecnabla   {\vec{\nabla}}
\def \vecB {\vec{B}}

\def \l {\left}
\def \r {\right} 

\maketitle

\begin{abstract}
We examine electron-transport coefficients  in magnetized hot and dense electron-ion plasma relevant in  binary neutron star merger simulation.   We calculate electrical and thermal conductivities in low density, high temperature, highly  magnetized plasma  of  binary neutron star mergers where quantum oscillatory behavior of electrons emerge. For pronounced thermodynamic effects, we consider zeroth Landau level population of electrons for the calculation of conductivity. We solve Boltzmann equation in presence of magnetic field to obtain the dissipative components of electrical and thermal conductivities. The dissipative coefficients are formulated considering frequency dependent dynamical screening   in the quantized electron-ion scattering rate. Numerical estimations show that the effect of  dynamical screening of photon propagator  on electrical and thermal conductivities is pronounced. We  observe that dynamical screening reduces the maxima of both the electrical and thermal conductivities by factors of thirty one and twenty  respectively leading to a reduction in the corresponding  time scales  of these coefficients.
The common scaling factor between electrical and thermal conductivity  is also observed to follow cubic relationship with  temperature violating Wiedemann-Franz law.
\end{abstract}

\section{Introduction}
Binary neutron star mergers and collider experiments  are the sources of most extreme  states of matter in the universe with densities several times nuclear saturation density  and temperatures  upto hundreds of MeV \cite{Dexheimer:2020zzs}. The recent detection of gravitational wave signal GW170817 originating from binary neutron star (BNS) merger by the LIGO and Virgo detectors have opened up a new era in multi-messenger astronomy \cite{LIGOScientific:2017vwq, LIGOScientific:2017zic, LIGOScientific:2017ync}. Additionally, short gamma-ray bursts (SGRBs) were also detected by the Fermi satellite GRB170817A indicating the presence of huge magnetic field in the merging event \cite{ Paschalidis:2016agf, Kawamura:2016nmk, Ruiz:2016rai}. These mergers are unique astrophysical objects of significant sources of gravitational radiation, electromagnetic as well as neutrino emission \cite{Palenzuela:2013hu}. They offer a novel avenue to study highly non-linear gravitational effects blended with complex micro-physical processes; serving as Einstein's richest natural laboratory \cite{Baiotti:2016qnr}.

In the event post merging, a remnant neutron star is created and if the remnant possess a mass beyond Tolman–Oppenheimer–Volkoff (TOV) limiting mass, the merged object survives for 10-100  milliseconds and collapses thereafter. The description of neutron star mergers requires the knowledge of General Relativistic Magneto Hydro-dynamics (GRMHD)\cite{Anderson:2008zp,Palenzuela:2008sf,Liu:2008xy,Dionysopoulou:2012zv,Palenzuela:2013kra,Kiuchi:2015qua,Dionysopoulou:2015tda,Kiuchi:2017zzg,Ruiz:2017due}.  Most of these general-relativistic simulations account for ideal Magneto Hydro-dynamics (MHD) limit. 
In a recent work Ref.\cite{Harutyunyan:2018mpe}, the authors  have pointed out that Hall effect  plays an important role in magnetic field decay of the merged object, hence,  the Hall effect needs to be incorporated in the ideal MHD merger simulation. In this Ref.\cite{Harutyunyan:2018mpe} the authors have considered  the electrical conductivity ($\sigma$) of warm neutron star crust in the non-quantizing scenario calculated in Ref.\cite{Harutyunyan:2016rxm}. Apart from these  studies, in Ref.\cite{Alford:2017rxf} authors have analysed the relevance of thermal conductivity ($\kappa$), viscous coefficients in BNS mergers in the non-magnetic scenario. In view of these recent studies \cite{Harutyunyan:2016rxm, Harutyunyan:2018mpe,Alford:2017rxf}, we analyse
 the importance of dissipative mechanism in the merger simulation  by evaluating quantized electrical and thermal conductivity coefficients with magnetically modified many-body effects in the present paper. We calculate quantized dissipative coefficients with frequency dependent  screening in both hot and dense plasma relevant in binary neutron star merger.   The results  can  be implemented   in analysing the  magnetothermal evolution \cite{Aguilera:2007xk} of the merged compact object as well.

 We consider  fully ionized plasma of electrons and ions. 
Heat and charge in  this medium are transported by electrons. The dominant electron transport mechanism is scattering on ions in the liquid phase. In presence of extreme magnetic field (B) (B$ \sim 10^{16} $G) and density ($\rho$) $(\rho \sim 10^{12}$ g cm$^{-3})$, the classical
  description of electrons breaks down. Therefore, one should incorporate Landau quantization of energy levels in the formalism.  This quantization occurs for a particular set of temperature, density and magnetic field in case of neutron star. Thus, the inclusion of Landau quantization eventually modifies the non-magnetic electrical conductivity to great extent\cite{Yakovlev1980, Hernquist1984, Potekhin:1996zh, Potekhin:1996hu}. In the present paper, we focus mainly on the strongly quantizing case, since in this domain, the transport coefficients receive major modification due to the magnetic field. 
      
 The calculation of electrical and thermal conductivities by solving Boltzmannn equation in ultra-compressed plasma have been studied by several authors over the last few decades \cite{1964Abrikosov, 1966ApJ...146..858H, Lampe:1968zz, 1976ApJ...206..218F}, see for a review \cite{Schmitt:2017efp}. This requires the information of scattering rate of plasma constituents. The calculations of scattering rates considering screened Coulomb potential have already been observed in different Refs.\cite{ Yakovlev1980, 1984MNRAS.209..511N, Hernquist1984,Potekhin:1996zh, Potekhin:1996hu}. In all these calculations it has been assumed that ions are static scatterers. This formulation can not be easily transported to the relativistic domain  where  dynamical effects are important for reliable description of transport coefficients. Medium modified  Hard-Thermal-Loop (HTL) propagators for hot  and Hard-Dense-Loop (HDL) for dense plasma  include dynamical effects of relativistic medium through frequency dependent screening \cite{Braaten:1989mz,Braaten:1990az,Altherr:1992mf,LeBellac:1996kr,Manuel:2000mk,Kalikotay:2020snc}. While Debye screening in plasma is related to the static/longitudinal photon exchange, the exchange of magnetic/ transverse photons contribute to dynamical frequency dependent screening of the plasma particles. 
 It is observed in different studies \cite{LeBellac:1996kr, Manuel:2000mk, Heiselberg:1993cr,Sarkar:2010bv,Sarkar:2012ww,Adhya:2012sq,Adhya:2013ima} that for ultra-degenerate case, both in Quantum Chromodynamics (QCD) and Quantum Electrodynamics (QED) plasmas, the transverse
interactions not only become important but they dominate over their longitudinal interaction.   
In a recent calculation \cite{Harutyunyan:2016rxm}, the authors have included many-body effects through the HTL modified propagator in the calculation of  non-quantized electrical conductivity in magnetized, warm neutron star crust.  Motivated by all these calculations  of dynamical screening in different  coefficients, we include HDL modified photon propagator in quantized $\sigma$ and $\kappa$ in the context of BNS merger in the present paper. We incorporate plasma screening through magnetic Debye mass. The current calculation is important in two ways. First, in this paper we  consider  the quantized transport coefficients in estimating dissipation coefficients in BNS merger. Second, this formulation includes Landau damping in quantized electron-ion interaction rate  in both hot and dense relativistic, magnetized plasma.  Here, we perform the calculations of $\sigma$ and $\kappa$ in an extreme scenario of  temperature  $\sim 12$ MeV, density $\sim 10^{12}$ g cm$^{-3}$ and magnetic field $ B\sim 10^{16} $ G. Finally, with the strongly quantized $\sigma$ and $\kappa$, we estimate the relevant dissipative time scales and compare it with the survival time period of the  post-merger object. Validity of Wiedemann-Franz law has also been discussed.

The paper is organised as follows. In section II, we derive the longitudinal electrical and thermal conductivities in a dynamically screened QED plasma.  Next, in section III  we present the constraints on temperature, magnetic field and density  of hot and dense plasma to become relativistic and strongly quantized in the BNS merger scenario.  We present numerical variations for $\sigma$ and $\kappa$ with temperature, magnetic field and density for the dense, relativistic plasma along with estimation of corresponding dissipative time scales in section III. Finally, we summarize and discuss the impact of dynamical screening on both the coefficients and decay time scales in section IV. Throughout the manuscript we will use following notation for four vectors $p=(\epsilon_p, p_z,  \vec{p}_{\perp})$ and  $k=(\epsilon_k, \vec k)$.  \footnote{We have used $c =k_B = \hslash = 1$.}  
\section{Electrical conduction from Transport Theory}
\label{formalism}
In the current paper, we consider fully ionized plasma of two components: electrons ($e$) and positive ions of charge 
$Ze$ ($Z$ atomic number of the nucleus). In  compact objects, the huge magnetic field quantizes the motion of electrons  in the QED plasma.
 In this section we derive the coefficients  for electrical conduction, electrical conductivity  and coefficients  for  thermal conduction, thermal conductivity,  in  magnetized electron-ion plasma from transport theory. In presence of magnetic field both the coefficients are anisotropic and the conductivity tensor is given below,

  \begin{equation}\label{sigmamatrix}
 \sigma/\kappa = \Bigg(\begin{matrix} 
 \sigma_{\perp}/\kappa_{\perp}& - \sigma_{H}/\kappa_{H}&0 \\ \sigma_{H}/\kappa_{H}&\sigma_{\perp}/\kappa_{\perp}&0\\0&0&\sigma_{\parallel}/\kappa_{\parallel}
 \end{matrix}
 \Bigg).
\end{equation}
\begin{figure}
\centering
\includegraphics[width=0.25\textwidth]{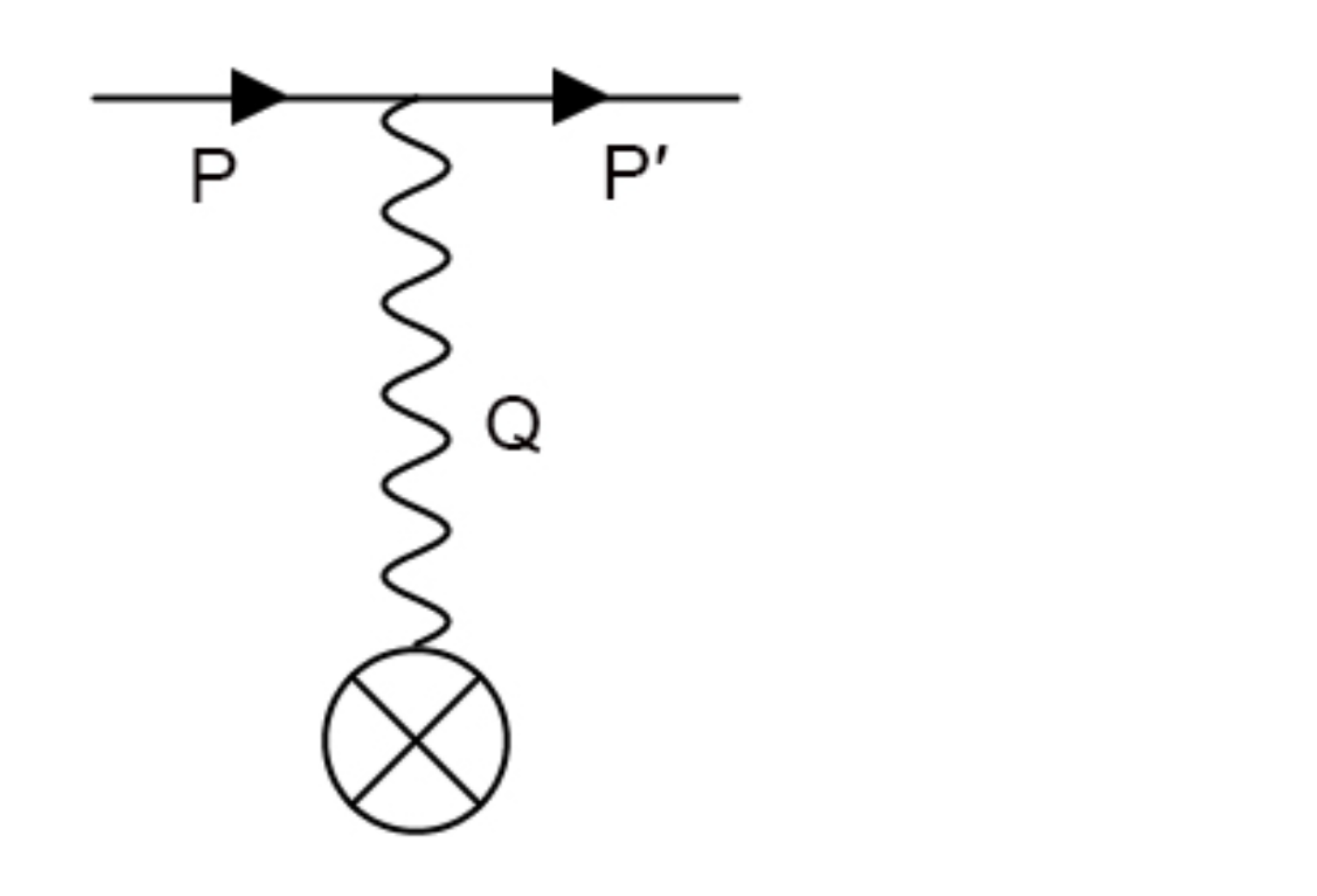}\includegraphics[width=0.22\textwidth]{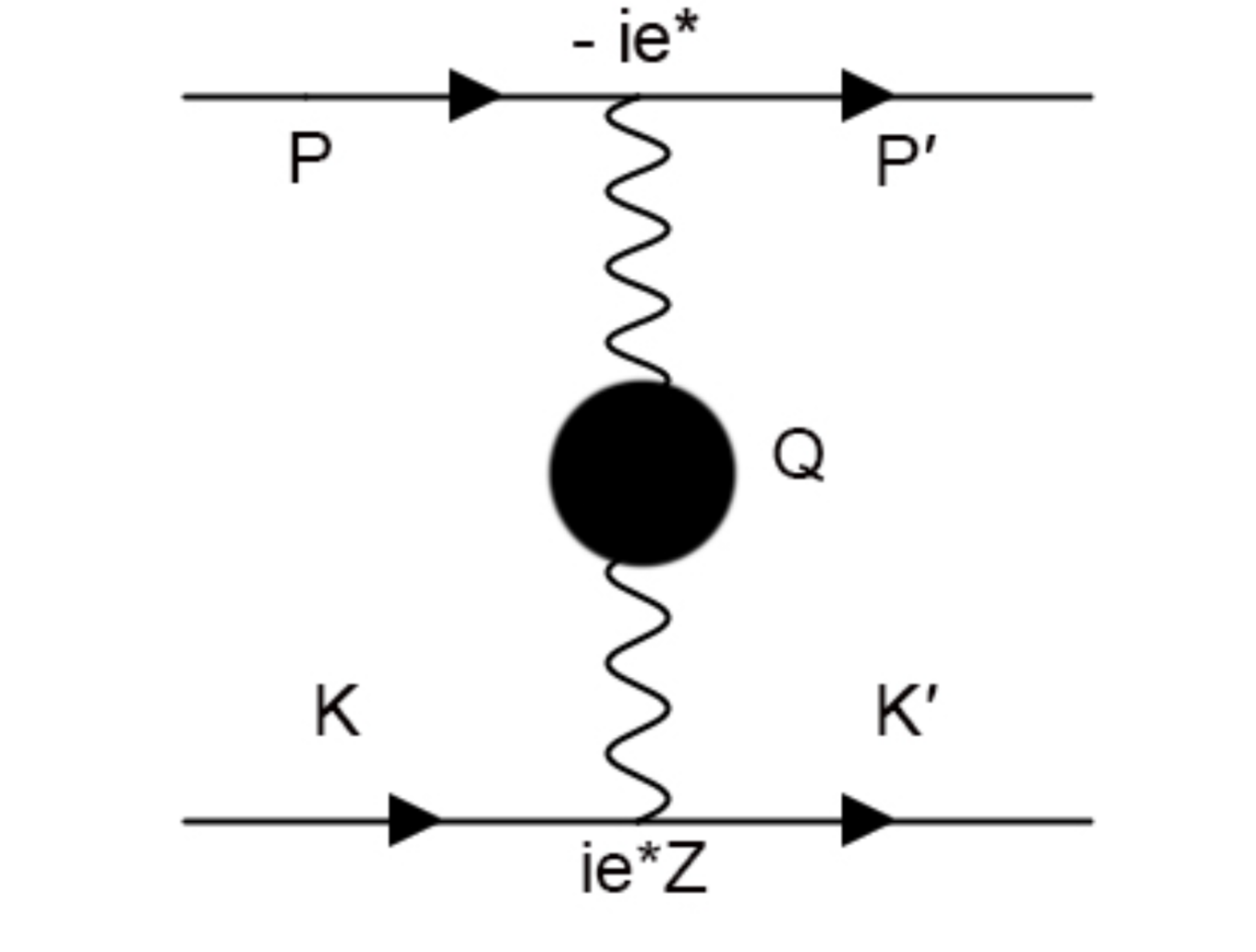}
\caption{Feynman diagram describing the electron-ion scattering amplitude in a static medium (left panel). Feynman diagram contributing to the electron-ion scattering amplitude via exchange of resummed HDL photon propagator  (right panel).}
\label{fig:feynman}       
\end{figure}
 In the above matrix equation, $\sigma_{\parallel}/\kappa_{\parallel}$ and $\sigma_{\perp}$/$\kappa_{\perp}$ are the parallel and perpendicular components of $\sigma$ and $\kappa$  respectively in presence of external magnetic field along $z$ direction.  $\sigma_{H}$/ $\kappa_{H}$,  are the Hall components of the conductivities. In the current paper we present the calculation of quantized $\sigma_{\parallel}/\kappa_{\parallel}$ in electron ion plasma. For the rest of the paper, we re-define $\sigma_{\parallel}$, $\kappa_{\parallel}$  as $\sigma$ and $\kappa$  respectively.

The $\sigma$ is related to the electric current density ($j$) and satisfies the constitutive relation  $j = \sigma E$ where $E$ is the electric field. For thermal conductivity $Q=-\kappa \nabla T$, where, Q is the amount of charge transported through the plasma under the temperature gradient $\nabla T$. $j$ and $Q$  are related to displacement of the electronic distributions from their equilibrium configuration due to the presence of electric field and temperature gradient in the plasma respectively and  can be calculated from kinetic theory,
 \begin{eqnarray}\label{ji}
 j&=&
 2e\int\frac{d^3 p}{(2\pi)^3} v_p\Phi\\
 Q&=&
 2\int\frac{d^3 p}{(2\pi)^3} (\epsilon_p-\mu)v_p\Phi.
\label{elec_current}
\end{eqnarray} 
In the above equation $e$ is the charge of an electron, $v_p$ is the velocity of the electrons, $\epsilon_p$ is the energy of electrons, $\mu$ is the chemical potential of electrons.   $\Phi$  is the off-equilibrium distribution function which arises due to the presence of electromagnetic field in plasma. $\Phi$ is obtained by solving Boltzmann equation in presence of magnetic field. In presence of  perturbation along $z$ direction, the distribution function ($f_{n, p_z, s}$) evolves according to the magnetically modified Boltzmann equation as given below \cite{Hernquist1984},
 \begin{eqnarray} \label{boltzmaneq}
\frac{\partial f_{n, p_z, s}}{\partial t} + v_z\frac{\partial f_{n, p_z, s}}{\partial z} - eE\frac{\partial f_{n, p_z, s}}{\partial p_z} =\mathcal{C}[f_{n, p_z, s}].
\label{be_b}
\end{eqnarray}
In the above equation, $ f_{n,p_z,s}$ describes the population of electrons defined by the quantum state $n, s, p_z$, $n$ is the number of the Landau level, $s$ is the spin and $p_z$ is the $z$ component of electron's momentum. 
$v_z$ is the $z$ component of the velocity of the particle.
The third term in the LHS of eq.(\ref{boltzmaneq}) arises from the Lorentz force term $\vec{F}=e(\vec E+\vec{v_{p}} \times \vec {B })$. In absence of external magnetic field, the Lorentz force term vanishes. 

The RHS of Eq.(\ref{be_b}) contains the information of scattering rate of electrons with the ions present in the medium,
\begin{eqnarray}\label{Cf}
\mathcal{C}[f_{n, p_z, s}] = \frac{\partial f_{n, p_z, s}}{\partial t}\Bigg|_{coll}= \sum_f I_{f i}\big(f_{n, p_z, s\rightarrow n', p'_z, s'} \big),
\end{eqnarray}
where, sum is over final state quantum numbers $n',p_z',s'$. $ I_{fi}$  is the electron-ion scattering rate from initial state ($i$) to the final state ($f$) in presence of $B$. $f_{n,p_z, s}$ is the initial state distribution function  and $f_{n',p_z',s'}$ is the scattered state distribution function.
 The distribution function of electrons has two parts in presence of electromagnetic field, 
\begin{eqnarray}\label{distfunc1}
f_{n, p_z, s} &=&f_{0}(\epsilon_p) + \Phi_{n,p_z,s},
\end{eqnarray}
 where, $f_{0}(\epsilon_p)$ is the equilibrium Fermi-distribution functions and $\Phi_{n,p_z,s}$ is the perturbation due to the electromagnetic field.

We now proceed to calculate the collision integral considering strongly quantizing magnetic field. To calculate the interaction rate, we consider an electron with momentum $p = (\epsilon_{p},p_z, \vec{p_{\perp}})$ and mass $m$ exchanges a virtual photon of momentum $q =$ ($q_0$,  $\vec q)$ with an in-medium ion of momentum $k = (\epsilon_{k},\vec k)$ and mass $M$. The electron emerges with momentum $p' = (\epsilon_{p'},p'_z, \vec{p'_{\perp}})$ and ion with momentum $k' = (\epsilon_{k'},\vec{k'})$ (Fig.(\ref{fig:feynman})). 
In order to obtain finite interaction rate, we use the HDL re-summed photon propagator with transverse and longitudinal components.

We start by re-visiting the expression of the interaction rate ($I_{fi}$) without electromagnetic modification \cite{Hernquist1984},
\begin{eqnarray}
\label{Ifi1}
&&I_{fi}=\frac{1}{2\epsilon_p}\int\frac{d^3p'}{(2\pi)^32\epsilon_{p'}}\int\frac{d^3k}{(2\pi)^32\epsilon_k}\int\frac{d^3k'}{(2\pi)^32\epsilon_{k'}}\nn\\
&&[f_{0}(\epsilon_p)g_{0}(\epsilon_k)\l(1-f_{0}(\epsilon_p')\r)-f_{0}(\epsilon_p')g_{0}(\epsilon_k')\l(1-f_{0}(\epsilon_p)\r) ]\nn\\
&&\l(2\pi \r)^4\delta\l(\epsilon_p+\epsilon_k-\epsilon_{p'}-\epsilon_{k'}\r)\delta^3\l(\vec {p}+\vec {k} -\vec {p'}-\vec {k'}\r)|{\cal M}_{fi}|^2\nn\\
\end{eqnarray}
where, ${\cal M}_{fi}$ is the electron-ion scattering matrix.   ${\cal M}_{fi}$ can be written as\cite{Harutyunyan:2016rxm},
\bea\label{eq:amplitude}
{\cal M}_{ fi}=-\Delta_LJ_0J'_0+\Delta_T
\bm J_t\bm J'_t=-{\cal M}_L+{\cal M}_T,
\label{mat_amp}
\eea
where, 
\bea\label{currents}
J^{\mu}&=&-e^*\bar{u}(p')\gamma^\mu u(p),\nn\\
J'^{\mu}&=&Ze^*v_k^{\mu}=Ze^*(1, \vec{k}/M), 
\eea
are the components of currents. $e^{\star}=\sqrt{4\pi } e$,  $v_k$ is the velocity of ion  with momentum $k$. The $\Delta_T$ and $\Delta_L$  are transverse and longitudinal HDL effective photon propagators respectively. The form of the electronic spinors ($u(p),  u(p')$) are given in the Appendix. A.

 To proceed further, here, we describe the screening mechanism of electron-ion plasma. In earlier calculations \cite{1964Abrikosov, 1966ApJ...146..858H, Lampe:1968zz, 1976ApJ...206..218F}, the authors have implemented  longitudinal component of photon propagator to  screen the Coulomb potential,
\begin{equation}
D_{\vec q} = \frac{1}{\vec q^{2}+m_{D}^{2}}.
\end{equation}
In the above equation $m_D^2= e^2dn_e/d\mu$, where, $n_e$ is the number density of electrons. Following linear response theory, there exists additional weak dynamical screening,  related to the
energy transfer to the constituents of plasma known
as Landau damping. This arises  because of non-zero frequency of the plasma. We implement the effects of non-zero frequency in both the electric and magnetic components of the photon propagator  computed within the HDL  formalism,
\bea\label{eq:Dmunu}
\Delta^{\mu \nu } (q)=
P^{\mu \nu }(q)\Delta_L
+ 
 Q^{\mu \nu }(q)\Delta_T,
\eea
where,   $P_{\mu\nu}$
and $Q_{\mu\nu}$ are the transverse and longitudinal projectors,
respectively, 
\bea\label{PQmunu}
P^{i j} (q) &=& -\delta^{ij} + \frac{q^i  q^j}{q^2},\\
Q^{00}(q) &=& -\frac{q^2}{|\vec q|^2} = 1 -\frac{q_0^2}{|\vec q|^2} = 1-y^2,
\eea
and the effective longitudinal ($\Delta_L$) and transverse ($\Delta_T$) propagators are given by the following expressions, 
\bea
\Delta_L=\frac{1}{q^2-\Pi_L},\nn\\
\Delta_T=\frac{1}{q^2-\Pi_T}.
\label{prop}
\eea
In the above equations $\Pi_T$ and $\Pi_L$ are the transverse  and longitudinal  HDL photon
self-energies and are given by,
\bea
\label{PiT}
\Pi_T (q) &=& 3m_D^2 \left[ \frac{y^2}{2} + \frac{y (1{-}y^2)}{2} 
\ln\left(\frac{y{+}1}{y{-}1}\right)\right],\nn\\
\Pi_L (q) &=& 3m_D^2\left[ 1-y^2-\frac{y(1-y^2)}{2}\ln\left(\frac{y{+}1}{y{-}1}\right)\right].
\eea

In presence of strong magnetic field, electron density present in the plasma changes, leading to a modification in the screening. At low temperature and strong magnetic field, the presence of sharp Fermi surface modifies the nature of screening. In the 
 relativistic domain $m\ll \mu$ , the Debye mass is given by \cite{Sharma:2010bx},
\begin{equation}
m_D^2 = 
\l(\frac{e}{\pi}\r)^2\l(\frac{eB}{2}\r). \label{eq:mDnrel}
\end{equation}	
\par

 One obtains eqn.(\ref{eq:mDnrel}) using $m_D^2= e^2dn_e/d\mu$, where, $n_e$ in the relativistic domain $(m\ll \mu)$ 
 is given by $eB\mu/2\pi^2$.  In the screening mass, we have ignored the finite temperature correction. This assumption is valid as we consider only the degenerate domain ($T<\mu$) of the plasma. 
In addition, the conductivities are weakly dependent on the screening mass as discussed in \cite{Hernquist1984}. Hence, inclusion of finite temperature Debye mass would  mark negligible correction in the final results of the conductivities.

In order to proceed further, we compute the phase space factor in the interaction rate given in eq.(\ref{Ifi1}). We  do not consider the change of momentum of ions in the phase space factor. Hence, the linearized phase space factor can be written as, 
\bea\label{phspacefac}
 &&\l[f_{n, p_z, s} g_{k} (1-f_{n',p'_z, s'})-f_{n',p'_z, s'}g_{k'}(1-f_{n,p_z, s})\r]\nn\\
 &&\simeq g_k\l(f_{n,p_z, s}-f_{n',p'_z, s'}\r).
\eea
The final expression for the interaction rate thus becomes (the details of the derivation are given in Appendix A), 
\bea\label{Ifi2}
I_{fi}&=&\frac{n_i }{2}\sum_{n',p_z',s'}\int  du\l(\Phi_{n,p_z,s}-\Phi_{n',p'_z, s'}\r)\nn\\
&&\l[\frac{1 }{3(u+\frac{\zeta}{3})(u+\zeta)}-\frac{v_k^2}{6u(u+\frac{\zeta}{3})}\r]{\cal F},
\eea
$n_i$ is the number density of ions and $u$, $\zeta$ and ${\cal F}$ are given in the Appendix A. In order to find the transport coefficients, it is useful to define a dimensionless scattering rate $a$ and a dimensionless perturbation to the distribution function ($\Psi$) defined as,
\bea\label{Iandf1}
&&\frac{I_{fi}}{n_iv_z\sigma_0}= a,\nn\\
&&\frac{eE}{\sigma_0n_i}\frac{\partial f_0}{\partial \epsilon_p}\Psi=\Phi,
\eea
where,  $\sigma_0=\pi Z^2 e^4/ \omega_B^2$ and $\omega_B=eB/m$. The electric charge $e$ is related to the fine structure constant by $\alpha = e^2/(4 \pi) = 1/137 $. Using the above two equations, we obtain the dimensionless form of the linearized Boltzmann equation as given below \cite{Yakovlev84},
\begin{equation}\label{eq:Beqdimless}
\sum_{n'~s'\gamma} a(n s \rightarrow n' s')\big(\Psi_{n' s'} - \gamma \Psi_{n s}  \big) = 1.
\end{equation}
$\gamma=\pm$ denotes the scattering channel for forward ($+$) and backward reactions ($-$). In the current paper, we present the results for the strongly quantizing scenario (i.e. zeroth Landau level) which provides the maximum effect with finite magnetic field in contrast to the  non-magnetic scenario. For zeroth Landau level, $n=n'=0$, spin degeneracy is absent and backward scattering ($\gamma=-1$) is the only allowed channel for scattering. Hence, after solving the dimensionless Boltzmann equation the off equilibrium distribution function is obtained as, 
  \bea\label{eq:phi}
 \Psi &=& \frac{E^{2} -1}{2Q_2} ,
 \eea
 where, $E=\epsilon_p/m$, $\Psi\equiv \Psi_{0,-1}$ and
\begin{eqnarray}
\label{eq:phi}
 &&Q_2 = \int_{0}^{\infty} e^{-u}\nn\\
 &&\times\left[\frac{2}{3\left(u+\frac{\zeta}{3}\right)\left(u+\zeta\right)}
 -\frac{v_k^2}{6u\left(u+\frac{\zeta}{3}\right)}\right] du.
\end{eqnarray}
 We obtain the expressions for $\sigma$ and $\kappa$ with only zeroth Landau level population by inserting the value of $\Phi$
 in the equations below, 
\bea
j&=&\frac{em\omega_B}{4\pi^2}\int_{m}^{\infty} \Phi d\epsilon_p,\nn\\
Q&=&\frac{m\omega_B}{4\pi^2}\int_{m}^{\infty} (\epsilon_p-\mu)\Phi d\epsilon_p.
\eea
We then insert $j$ and $q$ in the constitutive relations   $j=\sigma E$ and $Q=-\kappa \nabla T$ to obtain both $\sigma$ and $\kappa$. The final form of electrical conductivity is given below,
\begin{eqnarray}
\sigma&=&\frac{\delta_0}{\theta}
\int_{1}^{\infty}\Psi f_0(1-f_0) dE,
\label{sigma_final}
\end{eqnarray}
here, $\delta_0$ is a constant and is given by $\delta_0=m^4 b^2/8\pi^3 Z^2e^2n_i$,  $\theta= T/m $ and $b=B/B_c$, $B_c$ is the critical field given by $4.413\times 10^{13}$ G. $f_0(1-f_0)/T\equiv -\partial f_0/ \partial \epsilon_p$.

The  thermal conductivity coefficient can be obtained by integrating following expression,  
\begin{eqnarray}
\kappa&=&\frac{\pi^2  T}{3e^2}\Big[\frac{\delta_0}{\theta}\int_{1}^{\infty}\frac{(\epsilon_p-\mu)}{T^2}^2\Psi  f_0(1-f_0)dE\nn\\
&-&\frac{\delta_0}{T\theta}\frac{\int_{1}^{\infty}(\epsilon_p-\mu)\Psi  f_0(1-f_0)dE}{\int_{1}^{\infty}\Psi  f_0(1-f_0) dE}\Big].
\label{kappa_final}
\end{eqnarray}
 From the above two equations it is evident that at low temperature satisfying $(\epsilon_p-\mu)\sim T$, $\partial f_0/ \partial \epsilon_p=\delta(\epsilon_p-\mu)$. After integration thus $\sigma$ becomes temperature independent and $\kappa$ varies linearly with temperature $T$. Hence, $\kappa/\sigma$  varies linearly with temperature till $(\epsilon_p-\mu)\sim T$ is satisfied. This linear relationship of $\sigma/\kappa$ with temperature is known as  Wiedemann-Franz law.

In the next section, we  study the variation of $\sigma$ and $\kappa$ with different parameters and the effects of the frequency dependent screening derived in the current section.
\section{Results}
In this section we describe the behaviour of $\sigma$ and $\kappa$ with density, temperature, magnetic field in the hot and dense QED plasma. First we present the physical conditions of the plasma for the  calculation of the transport coefficients. 

\subsection{Physical conditions}
Physical properties of  the BNS merger, which 
forms an unstable configuration are different
from isolated neutron stars. We consider simplest possible constituents of post-merger object of electron-ion plasma with  fully ionized ions and  free mobile electrons in the low density (up to $10^{12} \mbox{g cm}^{-3}$), high magnetic field (up to $10^{17} \mbox{G}$) and high temperature ($T\sim 15$ MeV)  regime of BNS merger.  Electron density $n_e$ is related to ion density $n_i$ {\em via} $n_e=Zn_i$. We consider the magnetic field   present along the $z$ direction. Scattering of electrons with ions only contribute in electrical conductivity. 
 In the absence of magnetic field the electron density can be written as,
\begin{eqnarray}\label{elecdens1}
n_e&=&\frac{2}{(2\pi)^3}\int_0^{\infty}f_0\l(\epsilon_p \r) d^3p,
\end{eqnarray}

where, $f_0(\epsilon_p)=1/(exp(\frac{\epsilon_p-\mu}{T})+1)$,   $T$ is the temperature. In absence of magnetic field energy of the electrons are  given by $ \epsilon_p=\sqrt{ p_f^2+m^2}$, where, $p_f$ is the Fermi momentum. 

 The magnetically modified electronic energy states are obtained as solutions of Dirac equations in presence of finite magnetic field \cite{Akhiezer, Lifshitz}. The positive energy states are denoted by quantum numbers $\epsilon, p_z,n,s$.  $p_z$ is the electron momentum along the field which we consider along $z$ direction, $s =\pm 1$ is the helicity, and $n = 0,1,2$ enumerates the Landau levels. For non-zero  $B$, $\mu\equiv\mu(B)=(2\pi^2 n_e)/m\omega_B$.  The  energy of the relativistic electrons in presence of magnetic field is $ \epsilon_p = \sqrt{p_z^2 + m^2 + 2n \omega_B m}.$

 
 The ground Landau level is non-degenerate with respect to spin while the higher levels are doubly degenerate. 
 The number density of  electrons  in presence of magnetic field is written as,
 \begin{eqnarray}\label{elecdens2}
 n_e=\frac{m\omega_B}{(2\pi)^2}\int_{-\infty}^{\infty}dp_z \sum_{n,s}f(\epsilon_p),
 \end{eqnarray}

 where, the sums are over $n,s$. The magnetic field strongly quantizes the motion of electrons and different transport coefficients receive significant contribution when the electrons are confined to the zeroth Landau level. We do not consider the situation when ions receive quantum modifications due to the magnetic field.  Parameters which determine zeroth level population are as follows \cite{Chamel:2008ca},
 \begin{eqnarray}\label{TcerhoBpara}
&&T_{\mathrm{c}e}\approx
1.343\times 10^8\;B_{12}~\mathrm{K},\nn\\
&&\rho_B=7.045\times 10^3{A\over
Z}\;({B_{12}})^{3/2} \mathrm{\ g\ cm}^{-3}.
\label{temp_dens}
\end{eqnarray}
In the above equation, $B_{12}=B/10^{12}$, $ \omega_{ce}$ is the cyclotron frequency for electrons.  $B$ is strongly quantizing if $\rho<\rho_B$ and $T\ll T_{ce}$.

It is convenient to introduce the relativistic parameters $x_r=p_F/m \sim 1.008(\frac{\rho_6 Z}{A})^{1/3}$ (where $\rho_6=\rho/10^6$), $T_r\sim 5.930\times 10^{9}$ K. The electron-ion plasma is relativistic for $x_r\gg 1$ and $T\gg T_r$. 
 Thus the  electrons become relativistic when $T > 5\times 10^9$ K and density $\rho\sim 10^6$ $g$ $cm^{-3}$. 
 
 The momentum of an electron is related to the energy via the
relation $p_{nz}/m =\sqrt{(\epsilon_p/m)^2-2bn-1}$. From this expression one can obtain the
maximum Landau level that the electrons can populate and is given by the
integer part of $\nu=(E^2-1)/2b$. The energy of the electrons is constrained by the relation $(E^2-2b) < 1$ to meet the condition of lowest Landau level. This is an important condition   for the plots of $\sigma$ and $\kappa$ as we describe later
in this section. The parameters for density, temperature and magnetic field are appropriately chosen for relativistic quantized electrons to simultaneously meet the physical conditions applicable for the merging scenario also. We consider Fe and Mo for the numerical analysis of both the coefficients. The reason behind choosing these elements is given in the Appendix.B.
  \subsection{Variation with density}
\begin{figure*}
\centering
\begin{subfigure}[b]{1.0\textwidth}
\centering
\includegraphics[width=0.45\textwidth]{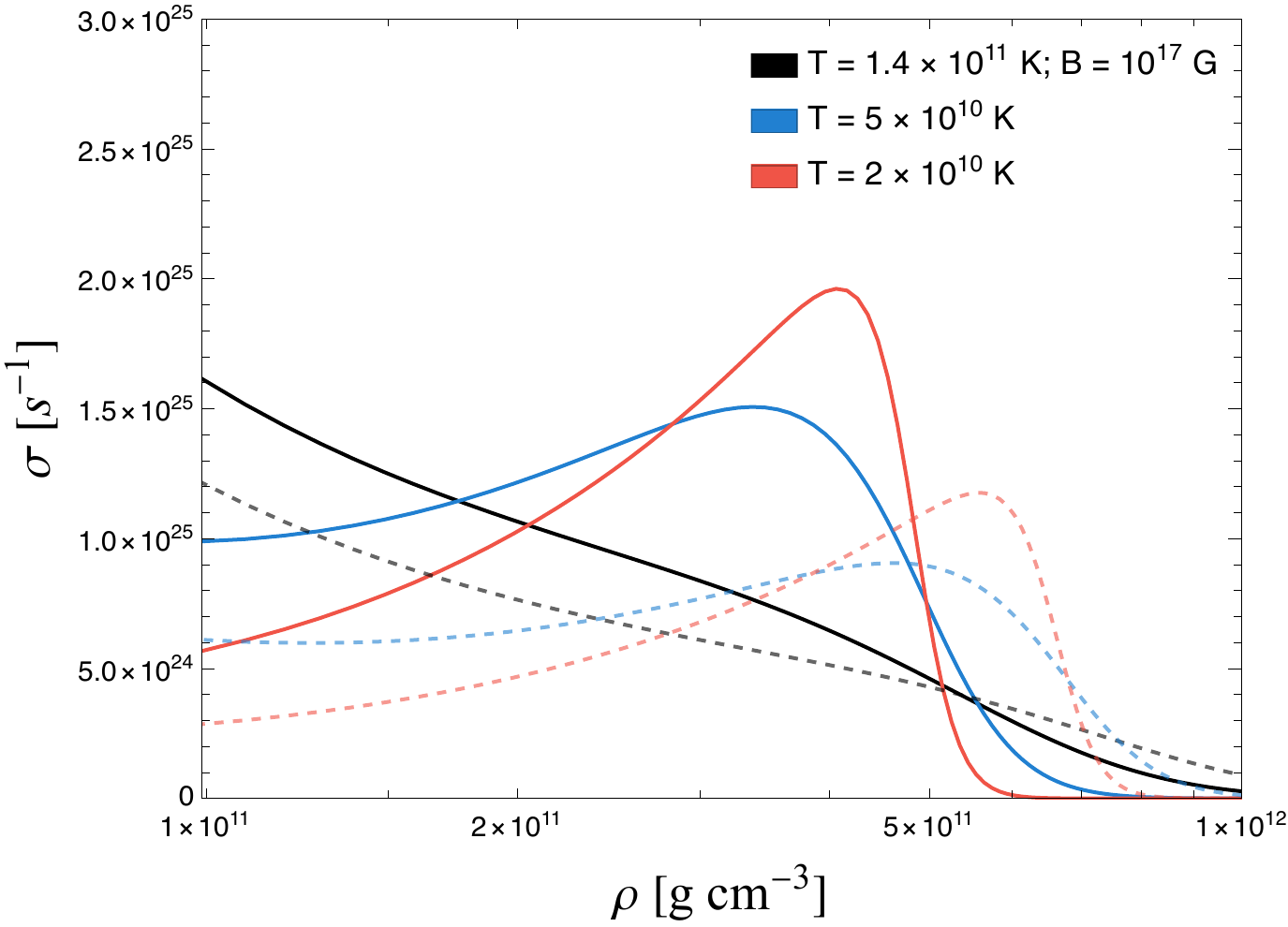}~~~~~\includegraphics[width=0.45\textwidth]{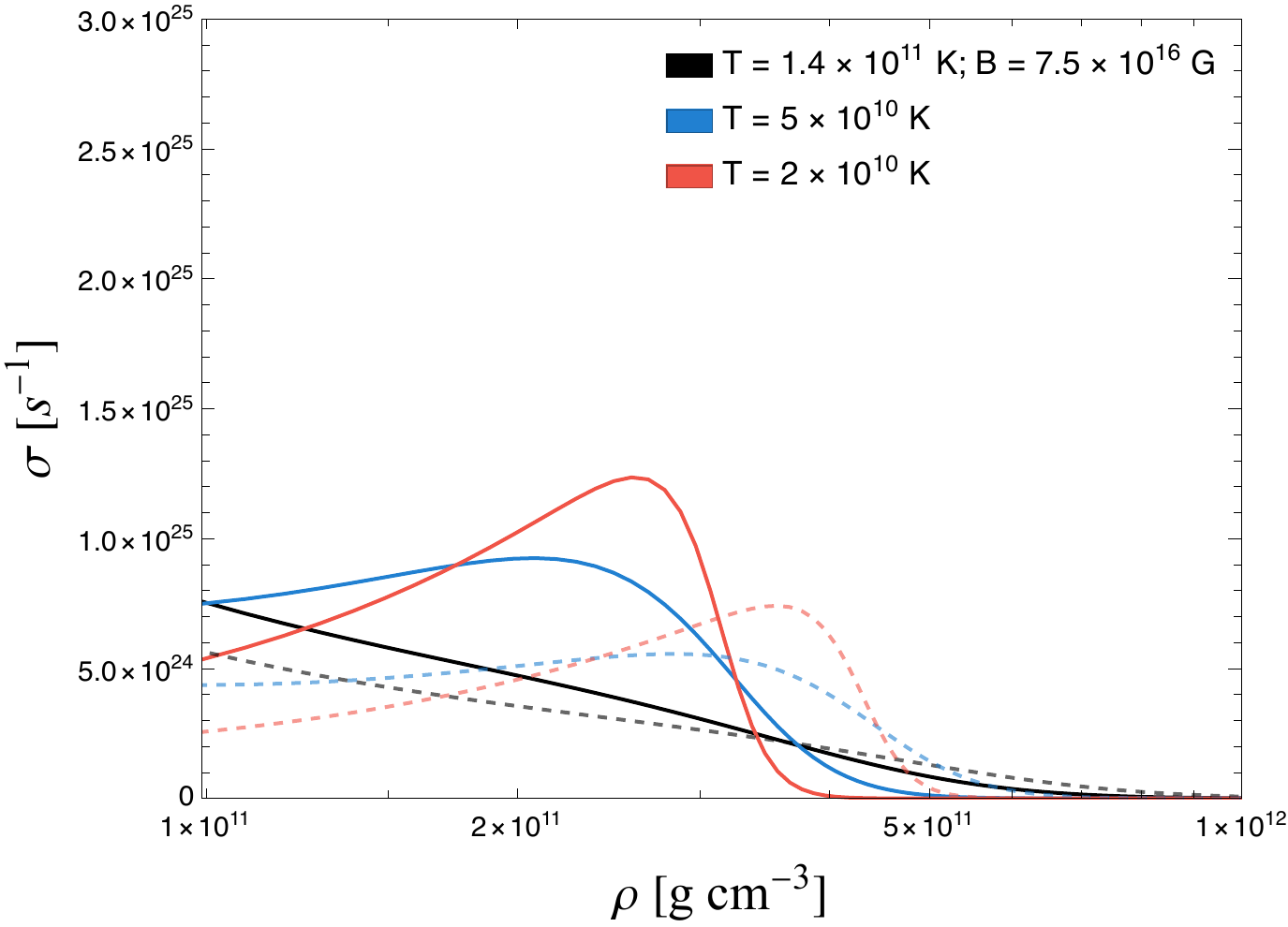}
\subcaption{Variation of $\sigma$ with $\rho$}
\label{fig:sigmavsrhocomb}
\end{subfigure}
\vskip 8pt
\hfill
\centering
\begin{subfigure}[b]{1.0\textwidth}
\centering
\includegraphics[width=0.45\textwidth]{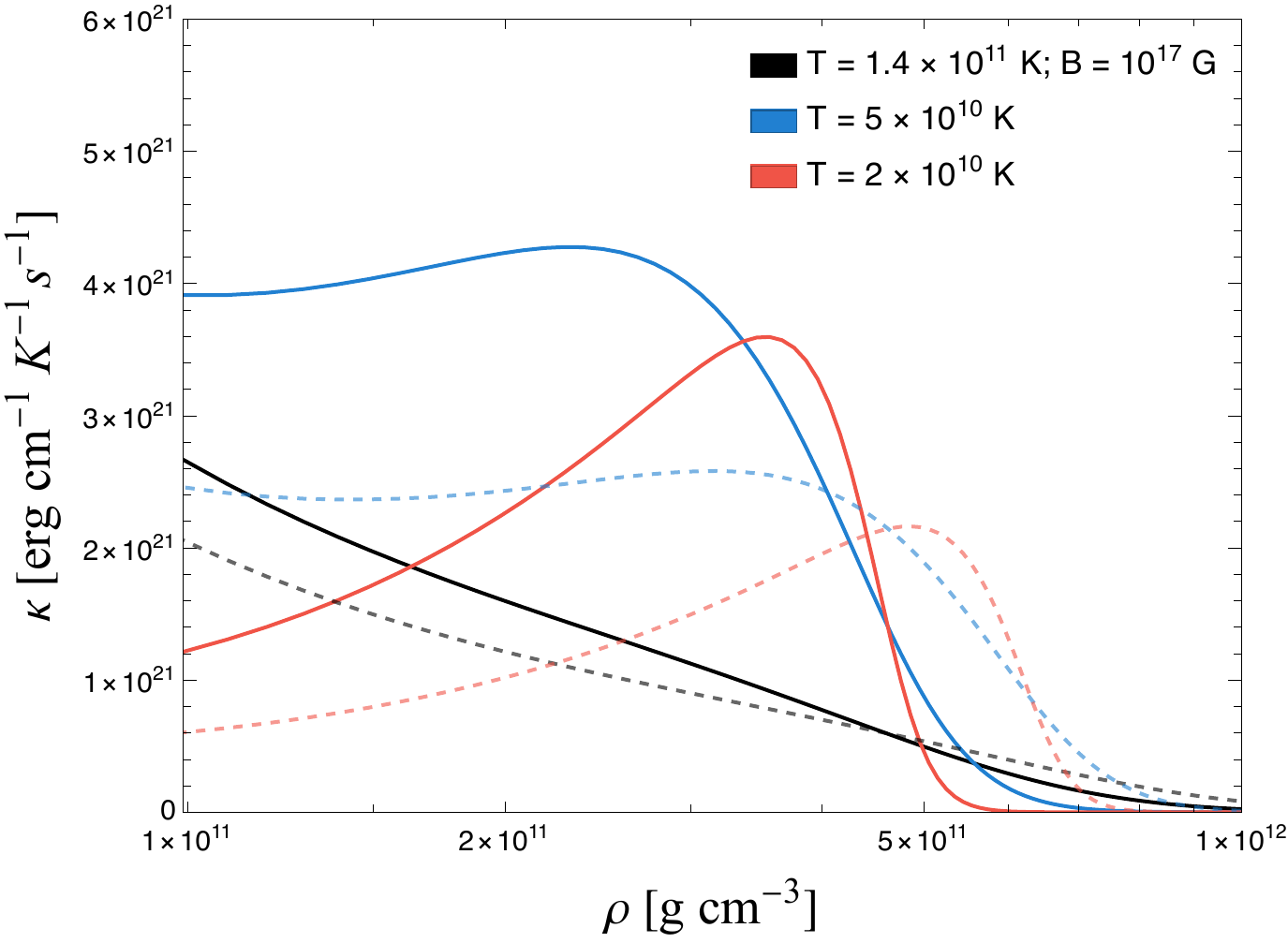}~~~~~\includegraphics[width=0.45\textwidth]{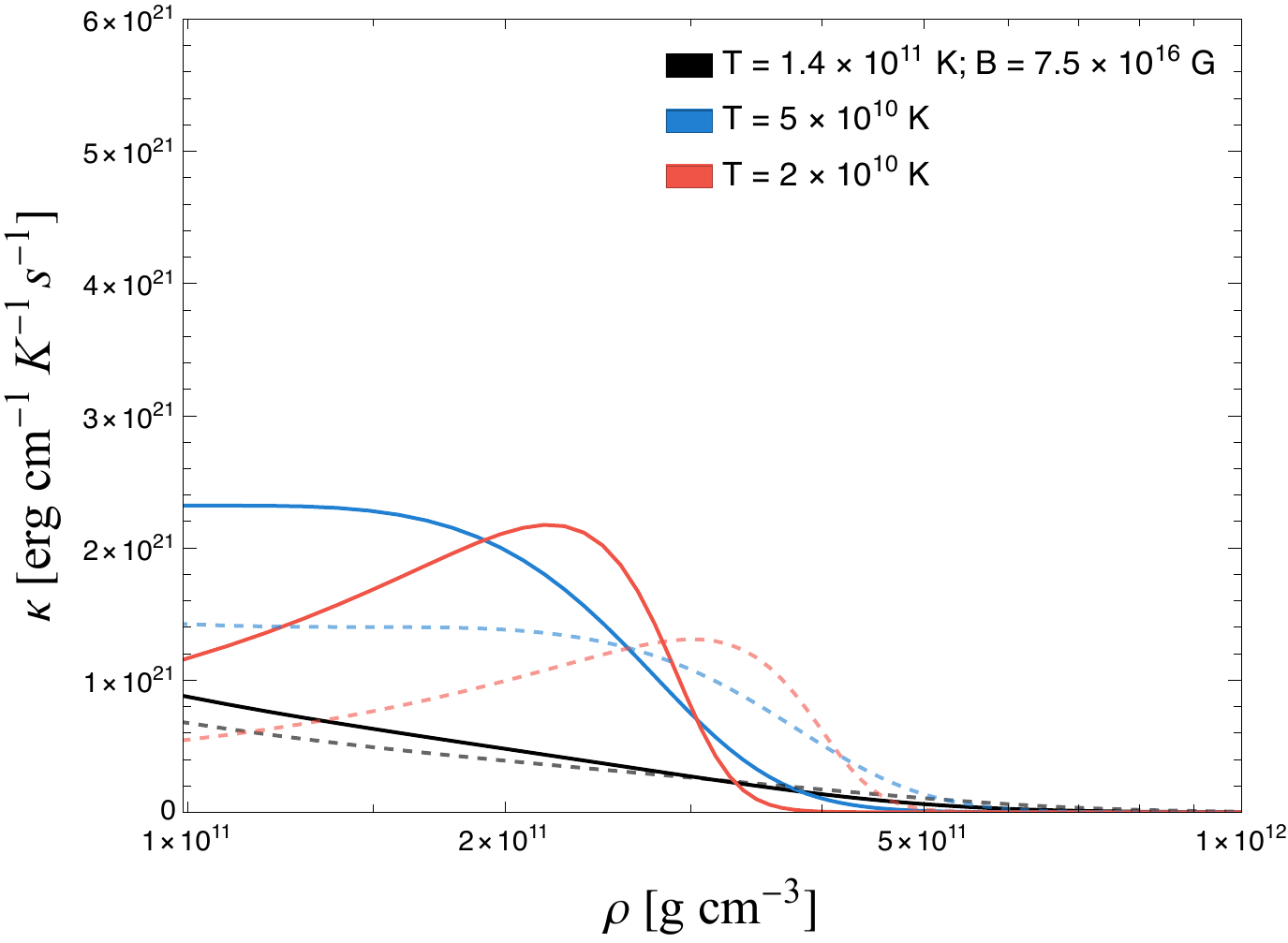}
\subcaption{Variation of $\kappa$ with $\rho$}
\label{fig:kappavsrhocomb}
\end{subfigure}
\caption{The comparison of $\sigma$ (upper plots) and $\kappa$ (lower plots) with $\rho$ for Fe (solid lines) and Mo (dotted lines). The left panel plots are for  magnetic field  $10^{17}$ G and the right ones are for $B=7.5\times 10^{16}$ G. In each plot we have three curves for each element corresponding to three different temperatures  $1.4 \times 10^{11}$ K,  $2 \times 10^{11} $ K,  $5 \times 10^{11} $ K. }
\label{fig:sigmakappavsrhocomb}
\end{figure*}

 Fig.\ref{fig:sigmakappavsrhocomb} shows the variation of $\sigma$ and $\kappa$ given in Eqs.(\ref{sigma_final}) and (\ref{kappa_final}) with $\rho$ for different temperatures and different magnetic fields for different elements Fe and Mo. In order to consider electrons to be relativistic, the density and temperatures are chosen as $\rho \gg 10^{6}$ $g$ $cm^{-3}$ and $T\gg 5\times 10^9$ $K$ respectively. For fixed $B$, $\mu$ increases with $\rho$ and electrons start to populate higher Landau levels. Since, we are interested in population of the zeroth Landau level, the density and temperature should also satisfy $\rho<\rho_B$ and $T\ll T_{ce}$ as given in Eq.(\ref{temp_dens}). 
 With  these two conditions, both the coefficients have been obtained by numerically integrating the expressions in Eqns. \ref{sigma_final} and \ref{kappa_final}.
   In fig.\ref{fig:sigmakappavsrhocomb}, the upper and lower panel plots are for the variation of $\sigma$ and $\kappa$ with $\rho$ respectively. The left panel  plots have been drawn considering magnetic field B=$10^{17}$ G and  $7.5\times 10^{16}$ G in the right panel. Each plot has three curves corresponding to three different temperatures $1.4 \times 10^{11} K$, $2 \times 10^{11} K$ and $5 \times 10^{11} K$ for each element. This is observed from the figure  that at temperatures $5 \times 10^{10}$ K and $2 \times 10^{10}$ K,  prominent humps are  present in both the coefficients. The origin of the hump is due to the fulfillment of the weak degeneracy condition ($|\epsilon_p-\mu|\sim T$) of electron distribution function. The nature of the curve resembles differentiated Fermi function at $T \ll \mu$. As the temperature increases, the hump gets flattened since electrons start becoming non-degenerate. In each plot of the figure curves are   drawn for two different materials Fe (solid lines) and Mo (dotted lines). 
  
\subsection{Variation with temperature and magnetic field}
\begin{figure*}
\centering
\begin{subfigure}[b]{1.0\textwidth}
\centering
\includegraphics[width=0.4\textwidth]{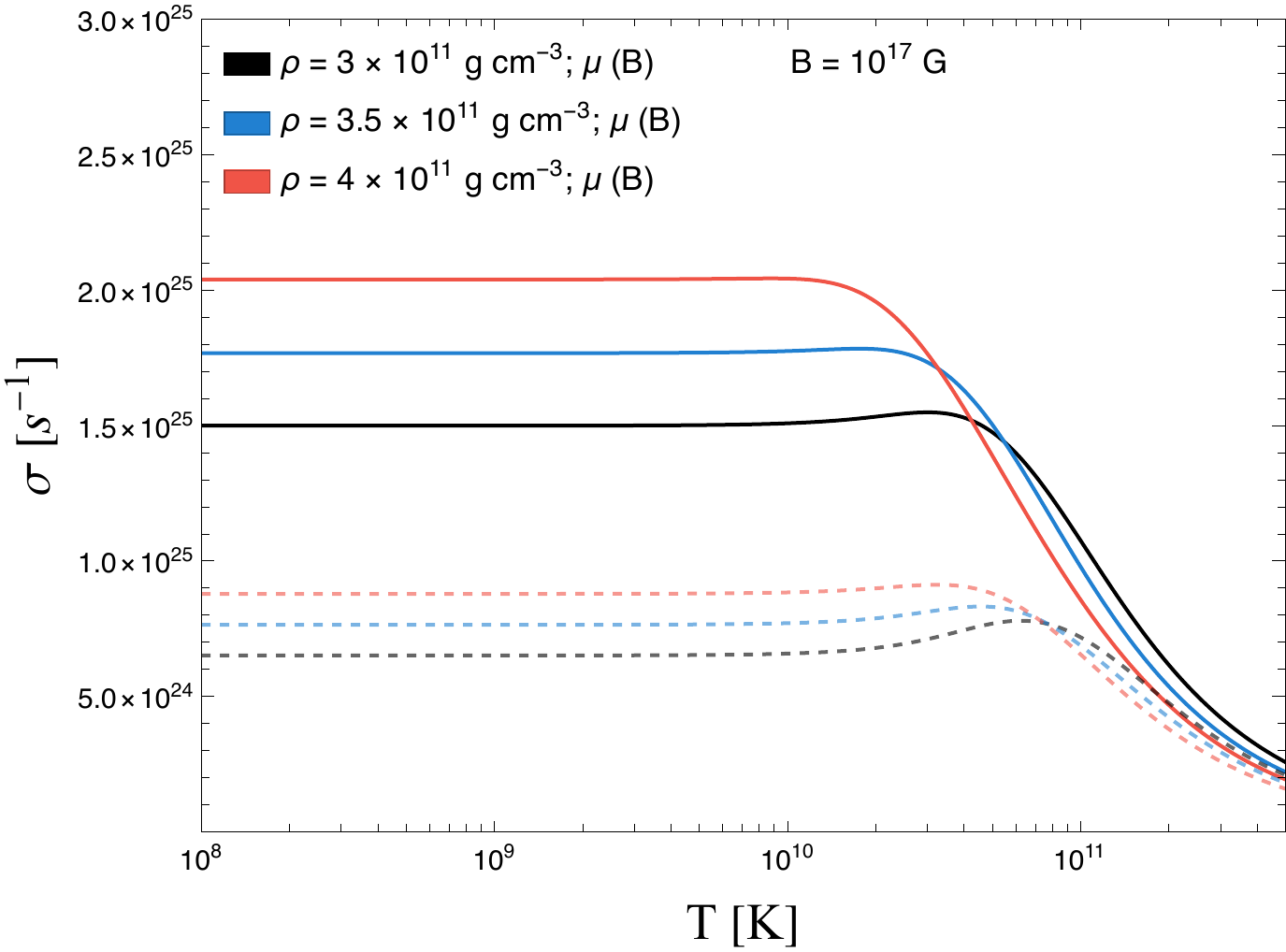}~~~~~\includegraphics[width=0.4\textwidth]{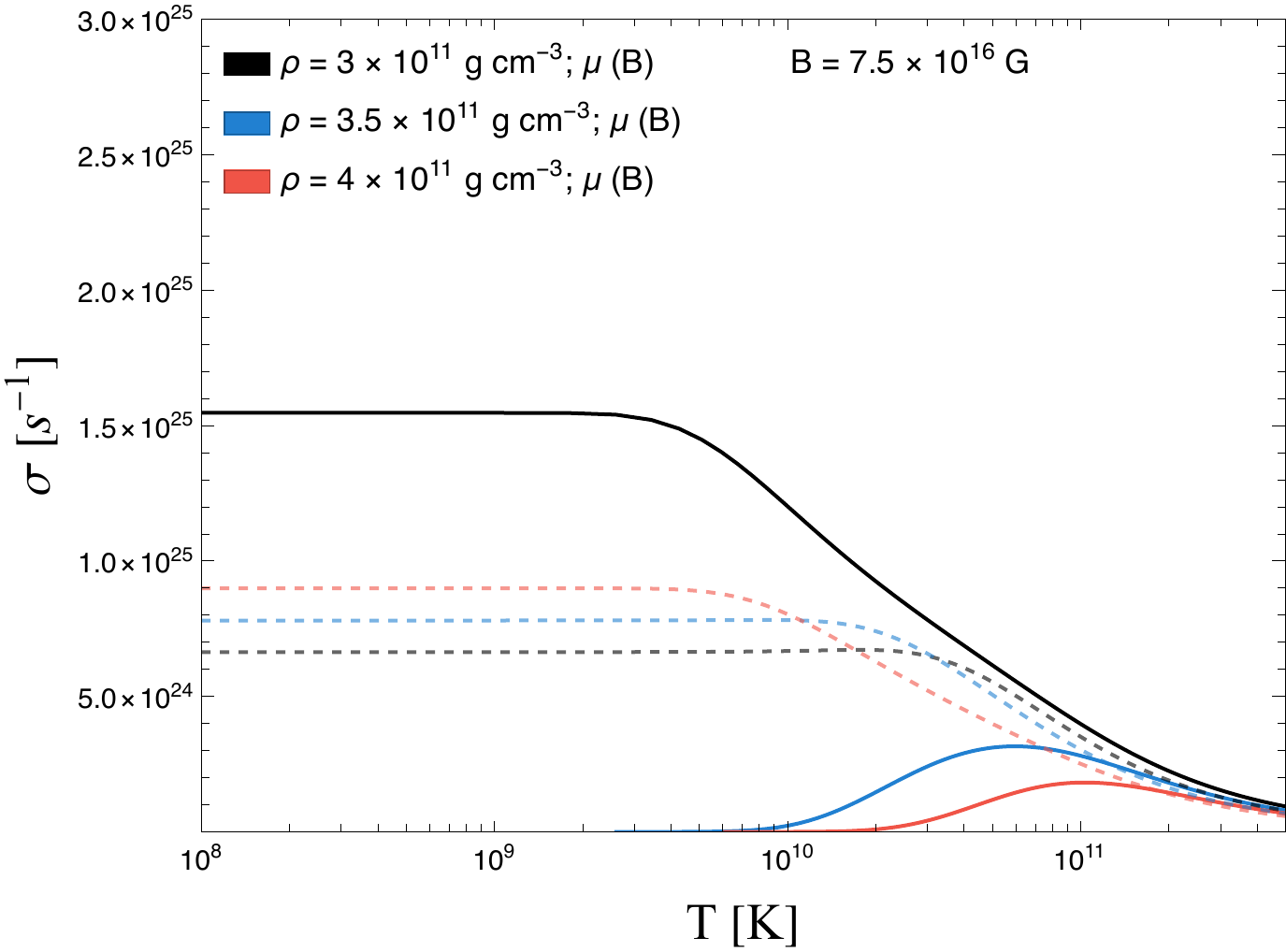}
\subcaption{Variation of $\sigma$ with $T$}
\label{fig:sigmavsTcomb}
\end{subfigure}
\vskip 8pt
\hfill
\centering
\begin{subfigure}[b]{1.0\textwidth}\label{fig:kappavsTcomb}
\centering
\includegraphics[width=0.4\textwidth]{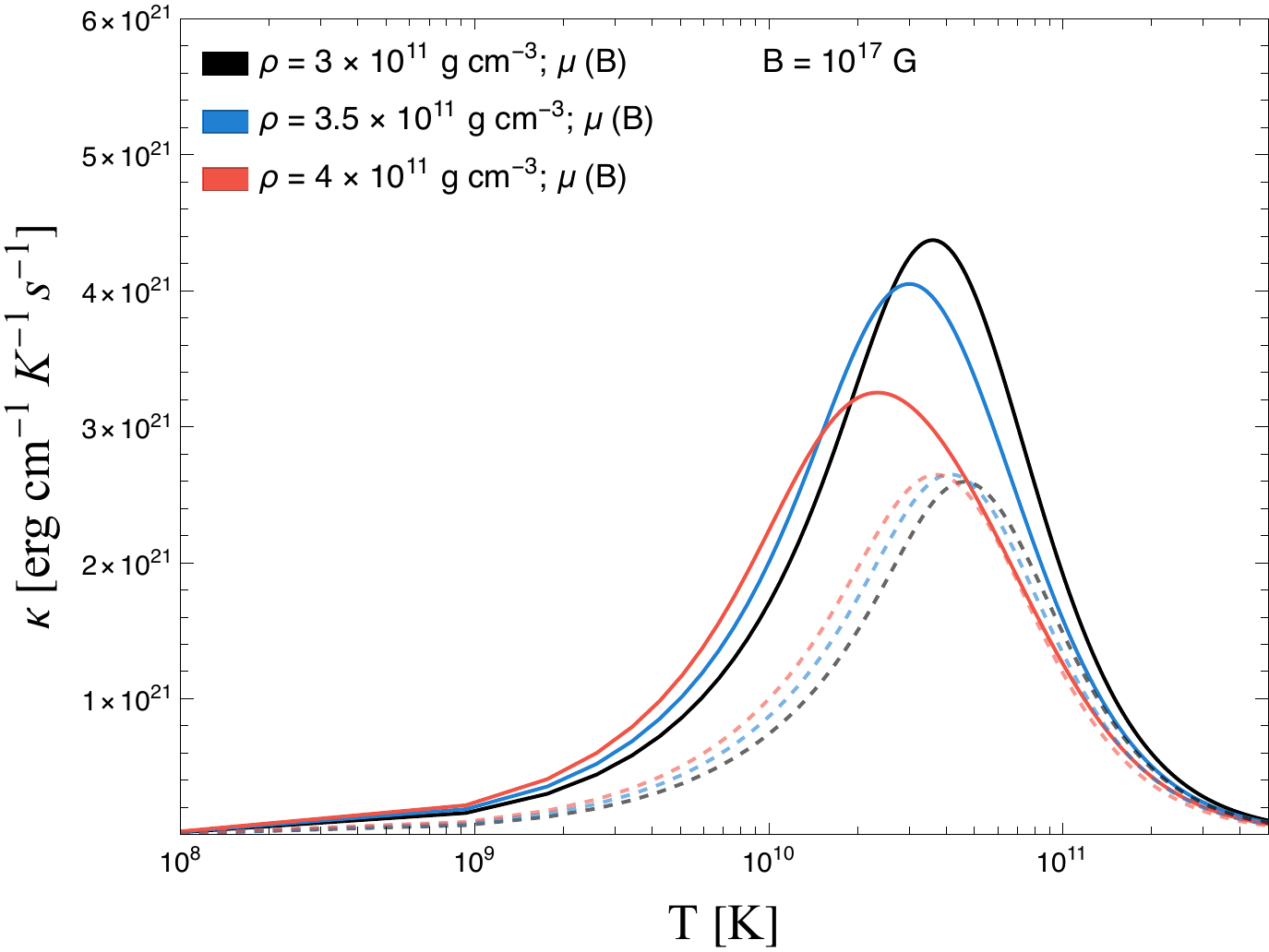}~~~~~\includegraphics[width=0.4\textwidth]{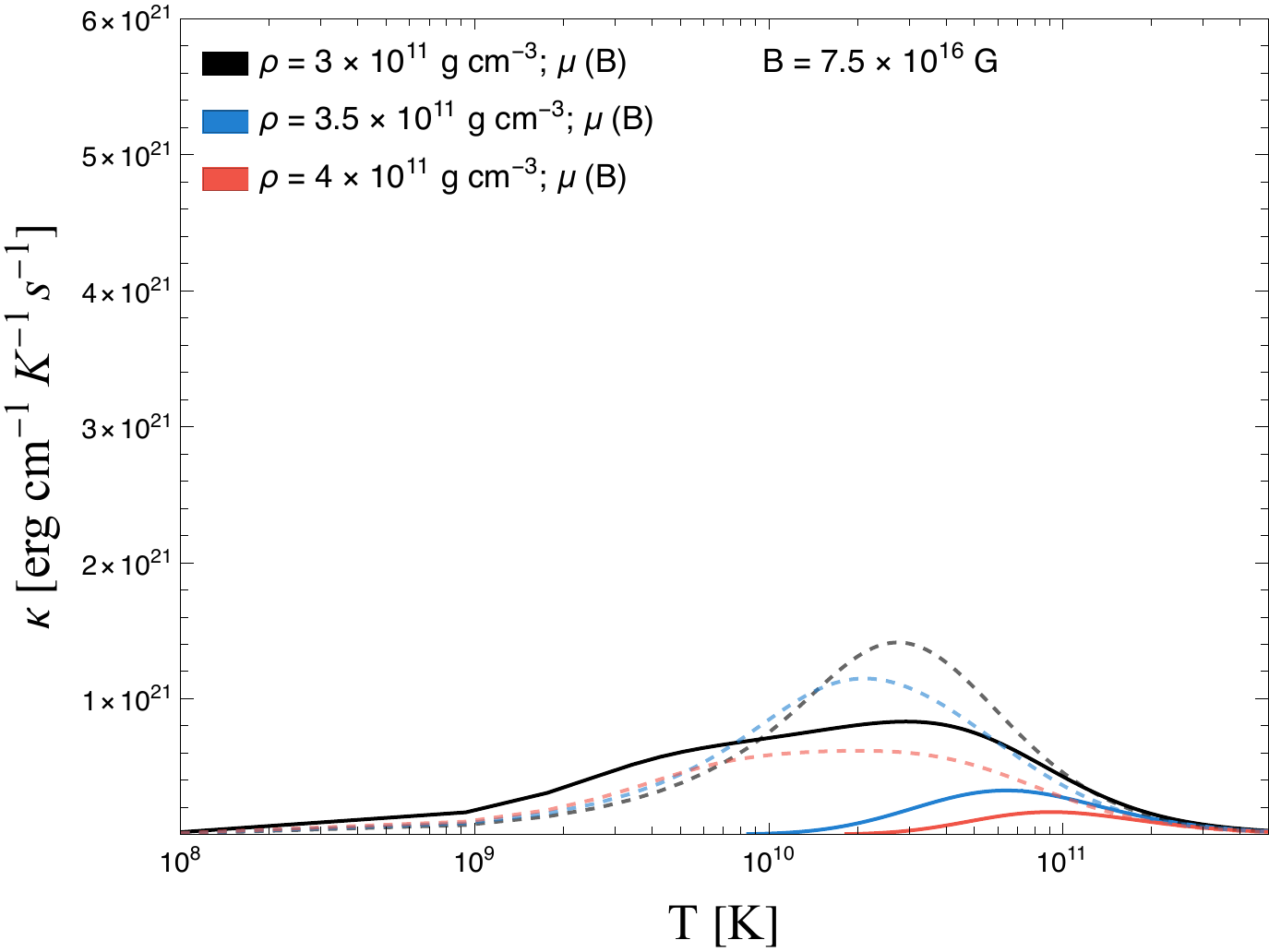}
\subcaption{Variation of $\kappa$ with $T$}
\end{subfigure}
\caption{The comparison of $\sigma$ (upper panel plots) and $\kappa$ (lower panel plots) with T for Fe (solid lines) and Mo (dotted lines). The left panel plots are for  magnetic field  $10^{17}$ G and the right ones are for $B=7.5\times 10^{16} $ G. In each plot we have three curves of each element corresponding to three different densities  $3\times 10^{11}$ gcm$^{-3}$,  $3.5 \times 10^{11} $ gcm$^{-3}$,  $4 \times 10^{11}$ gcm$^{-3}$. }
\label{fig:sigmakappavsTcomb}
\end{figure*}

Fig.(\ref{fig:sigmakappavsTcomb}) shows variation of $\sigma$ and $\kappa$ with $T$ for different densities. 
The $\sigma$ in fig.\ref{fig:sigmavsTcomb} can be fitted as $\sigma=(a+b\times T^c)^{-1}$,
with $a=1.45\times 10^{-25}$,
$b=1.05\times 10^{-46}$ and
$c=1.919$. At low temperature the effect of $T^c$ is very small, hence, $\sigma$ is constant. On the other hand at high temperature $\sigma\propto  T^{-c}$ and decreases with temperature. Thus, at higher temperatures, the electrons become classical obeying the inverse dependence of temperature. 

The temperature dependence of $\kappa$ can be attributed to low temperature behaviour of Fermi function whose derivative shows a hump when $|\epsilon_p-\mu|\sim T$. In each plot of fig.(\ref{fig:sigmakappavsTcomb}), we have three curves of each Fe (solid lines) and Mo (dotted lines) for three different densities  $3\times 10^{11}$ gcm$^{-3}$,  $3.5 \times 10^{11}$  gcm$^{-3}$ and $4 \times 10^{11}$  gcm$^{-3}$.

\begin{figure*}
\centering
\begin{subfigure}[b]{1.0\textwidth}\label{fig:sigmavsBcomb}
\centering
\includegraphics[width=0.4\textwidth]{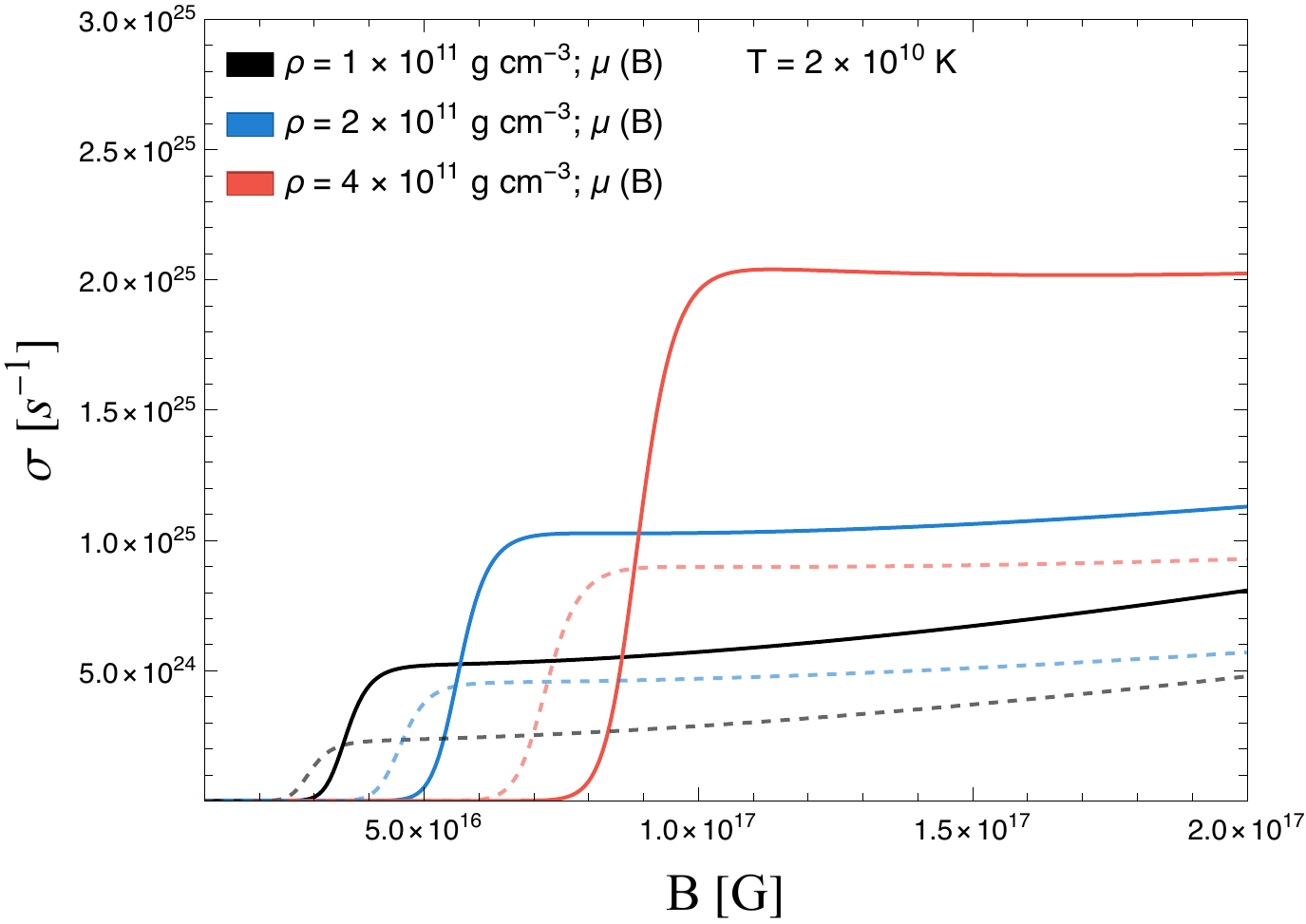}~~~~~\includegraphics[width=0.4\textwidth]{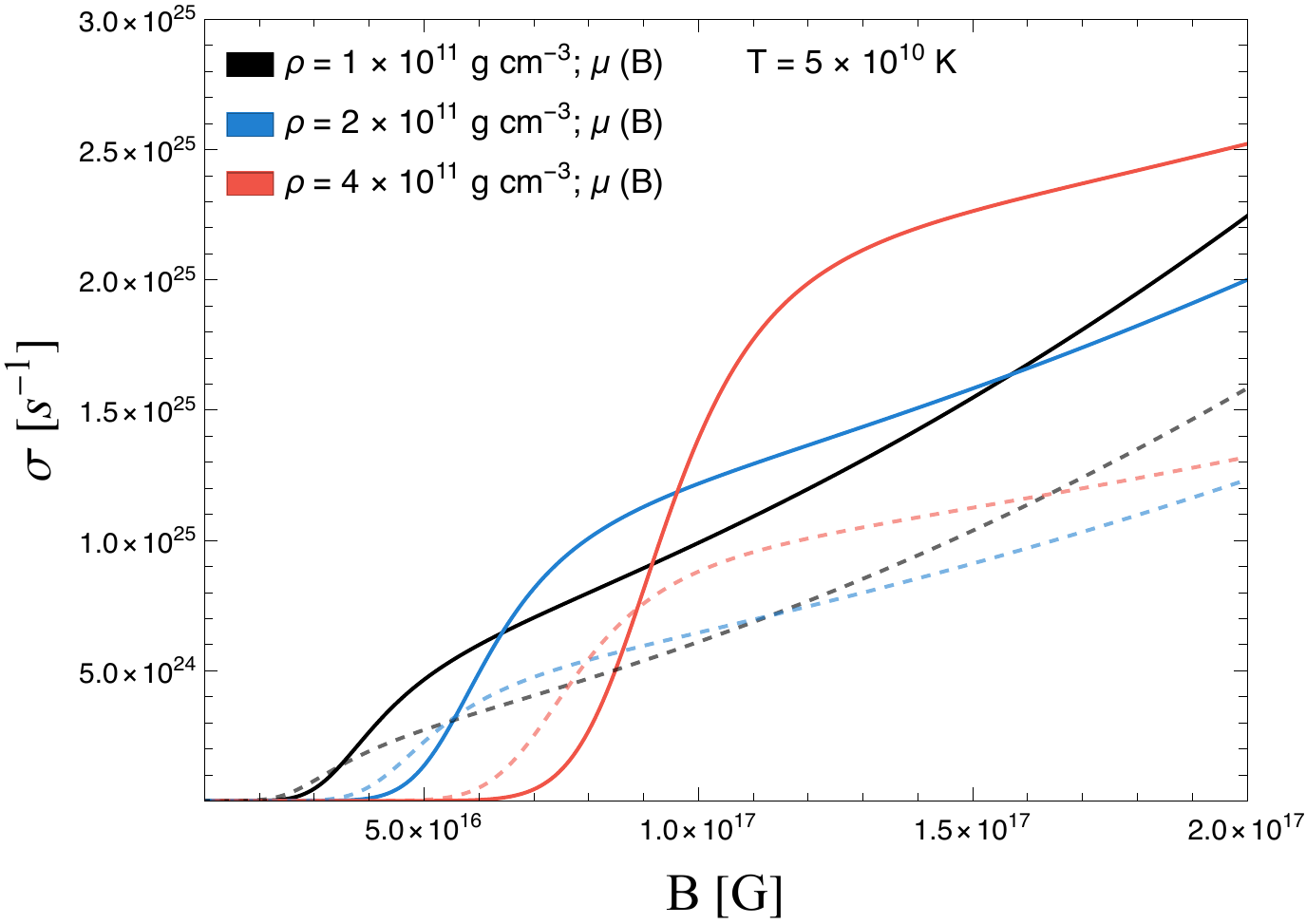}
\subcaption{Variation of $\sigma$ with $B$}
\end{subfigure}
\vskip 8pt
\hfill
\centering
\begin{subfigure}[b]{1.0\textwidth}\label{fig:kappavsBcomb}
\centering
\includegraphics[width=0.4\textwidth]{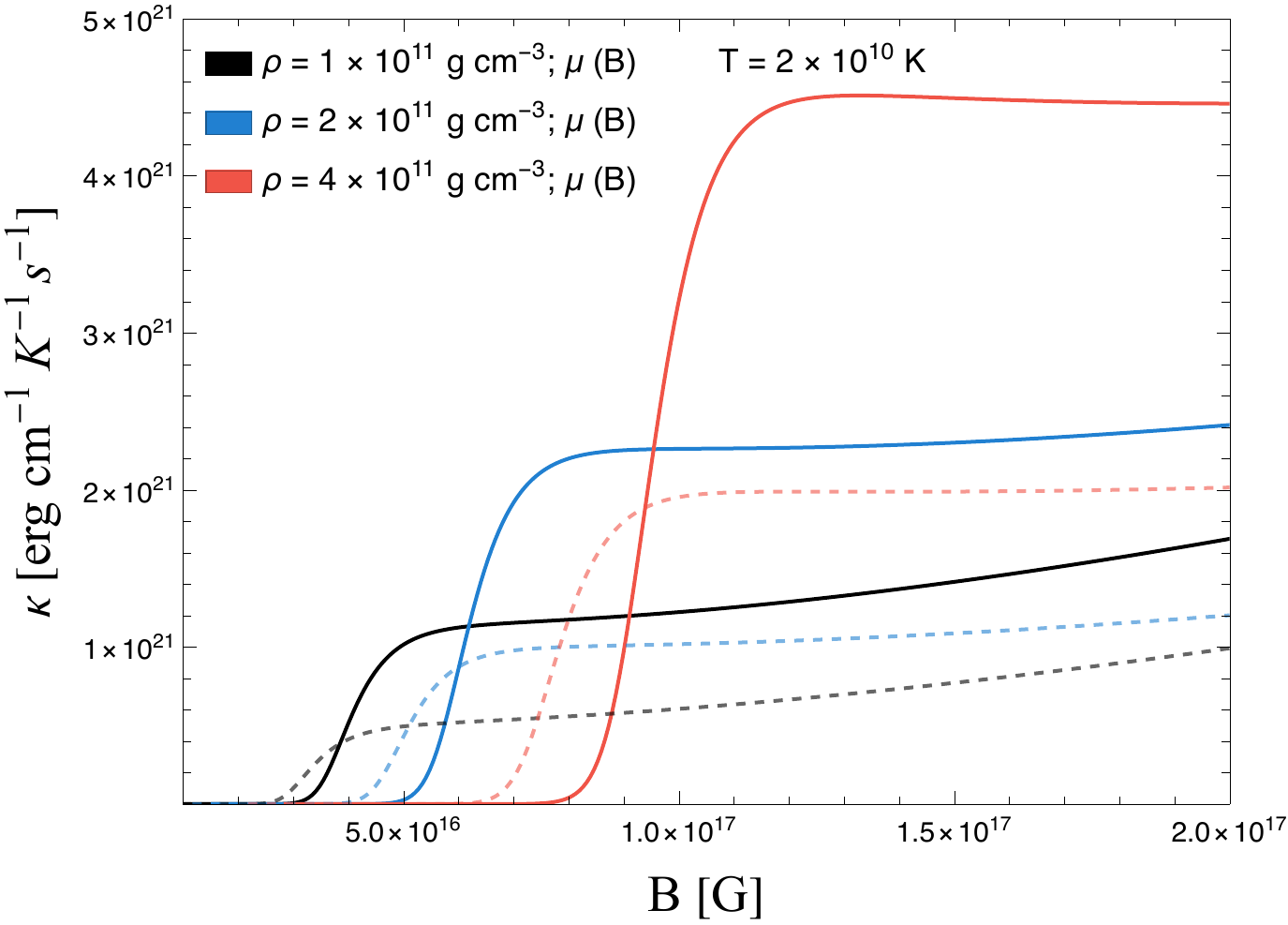}~~~~~\includegraphics[width=0.4\textwidth]{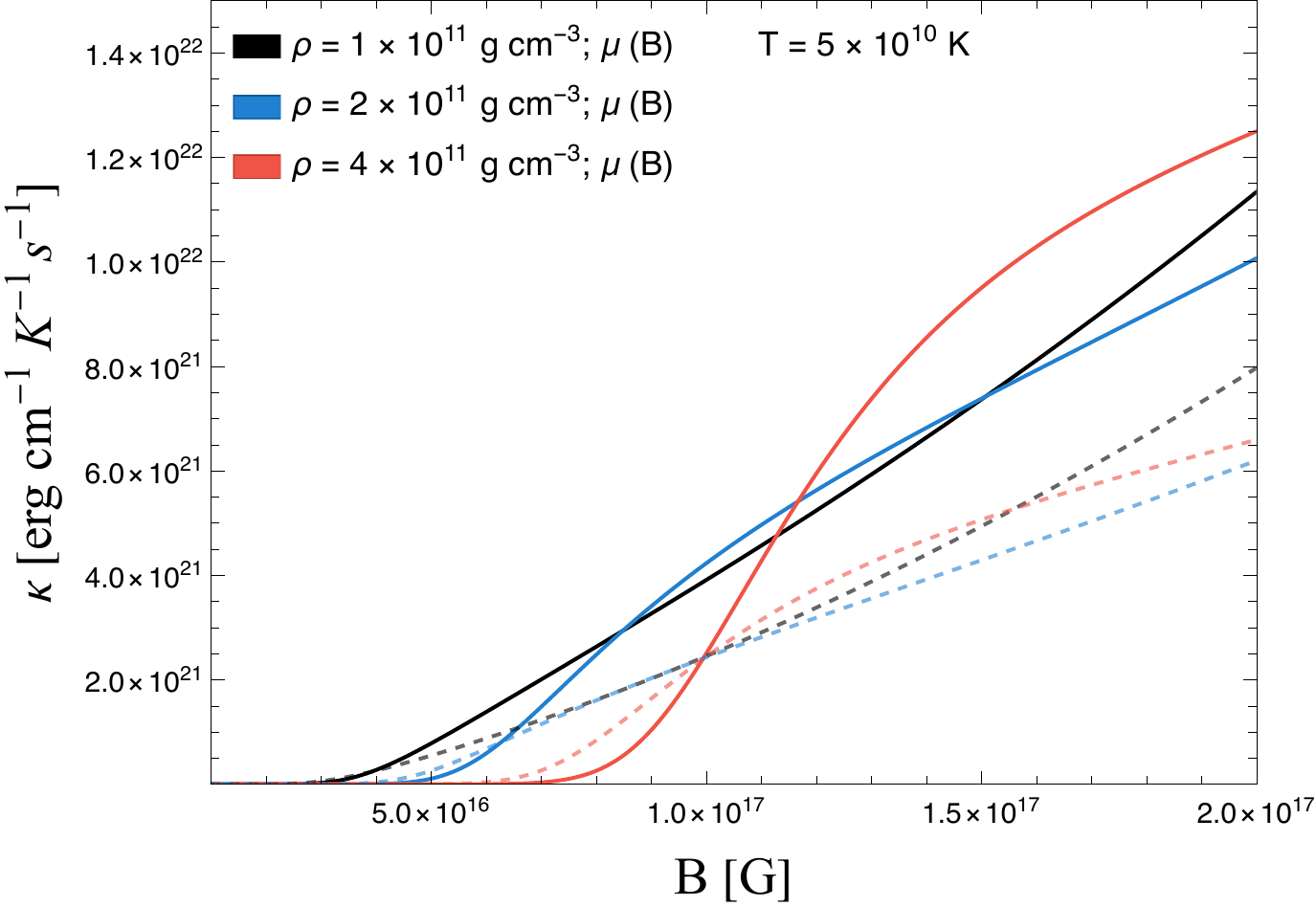}
\subcaption{Variation of $\kappa$ with $B$}
\end{subfigure}
\caption{The comparison of $\sigma$ (upper panel) and $\kappa$ (lower panel) with B for different densities. The temperatures chosen are $2\times10^{10} $ K (left panel) and $5\times10^{10} $K (right panel). The choice of elements are Fe  (solid lines) and Mo (dotted lines).}
\label{fig:sigmakappavsBcomb}
\end{figure*}

Fig.(\ref{fig:sigmakappavsBcomb}) shows the variation of  $\sigma$ and $\kappa$ with $B$ for different densities. The fitting parameters for the $\sigma$ with $B$ is given by,
$\sigma=5.15\times 10^{27}-2.71\times 10^{11}B+5.30\times 10^{-6}B^2$. Increasing the magnetic field, $\sigma$  increase with $B$ and saturates.  Variation of $\kappa$ with $B$ 
 is given by  the polynomial equation $\kappa=2.43\times 10^{23}-81.2\times 10^5B+8.85\times10^{-11}B^2-3.13\times 10^{-28}B^3$.
 $\kappa$ shows similar trend like $\sigma$ with $B$.
\vskip 12pt

\vskip 12pt
\vskip 12pt
\begin{figure*}
\centering
\begin{subfigure}[b]{1.0\textwidth}\label{fig:sigmarelnonrel}
\centering
\includegraphics[width=0.4\textwidth]{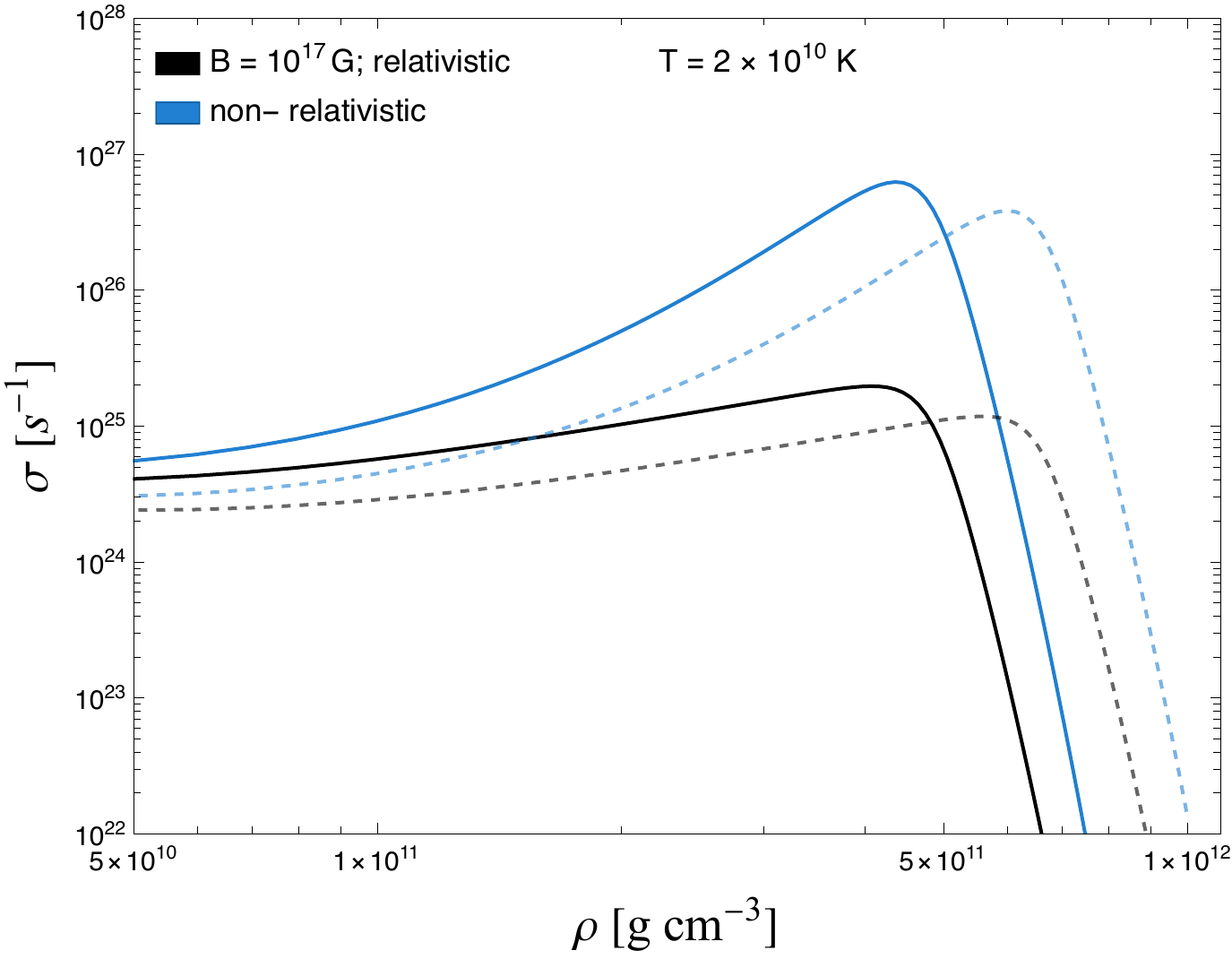}~~~~~\includegraphics[width=0.4\textwidth]{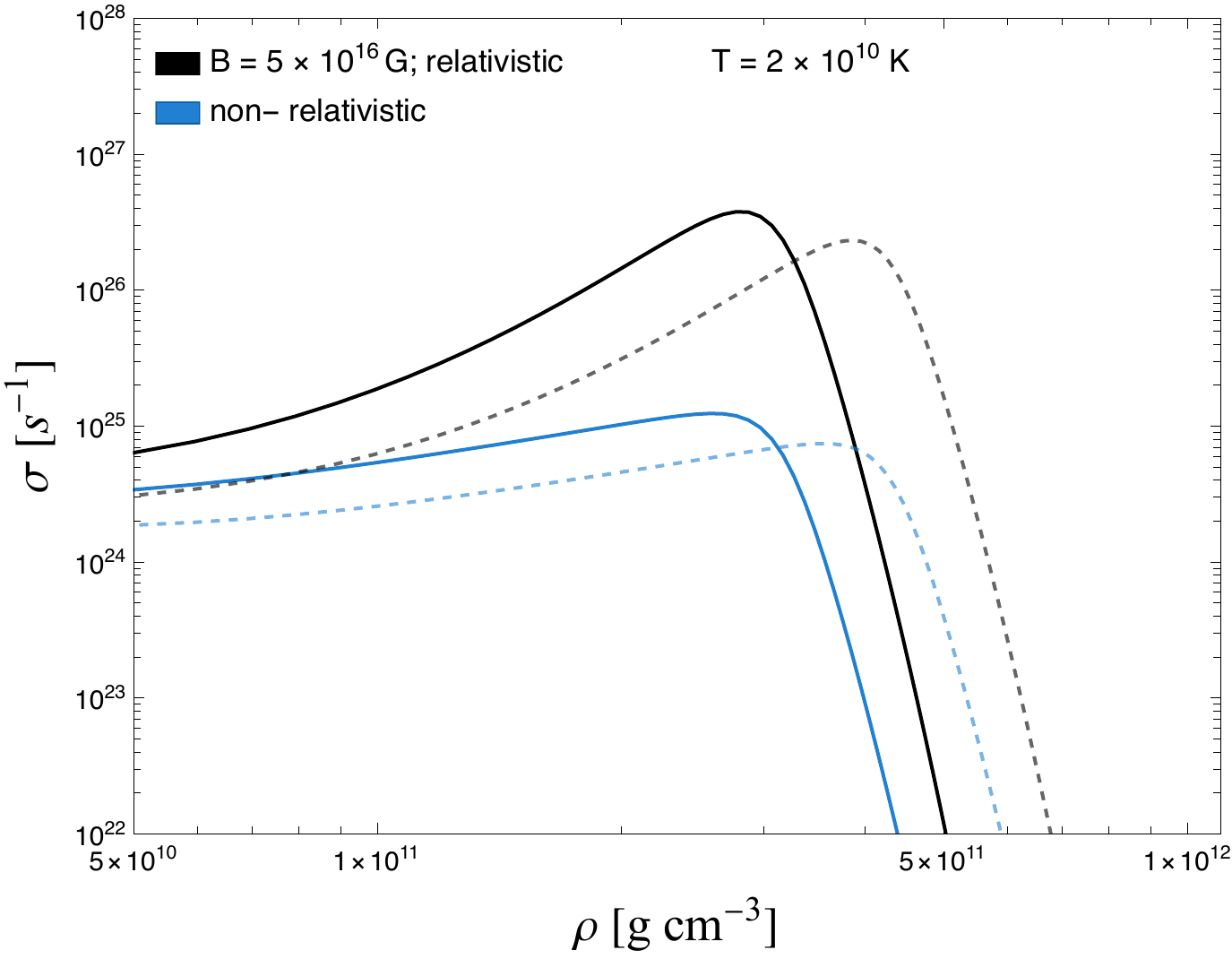}
\subcaption{Effect of HDL propagator on $\sigma$ }
\end{subfigure}
\vskip 8pt
\hfill
\centering
\begin{subfigure}[b]{1.0\textwidth}\label{fig:kapparelnonrel}
\centering
\includegraphics[width=0.4\textwidth]{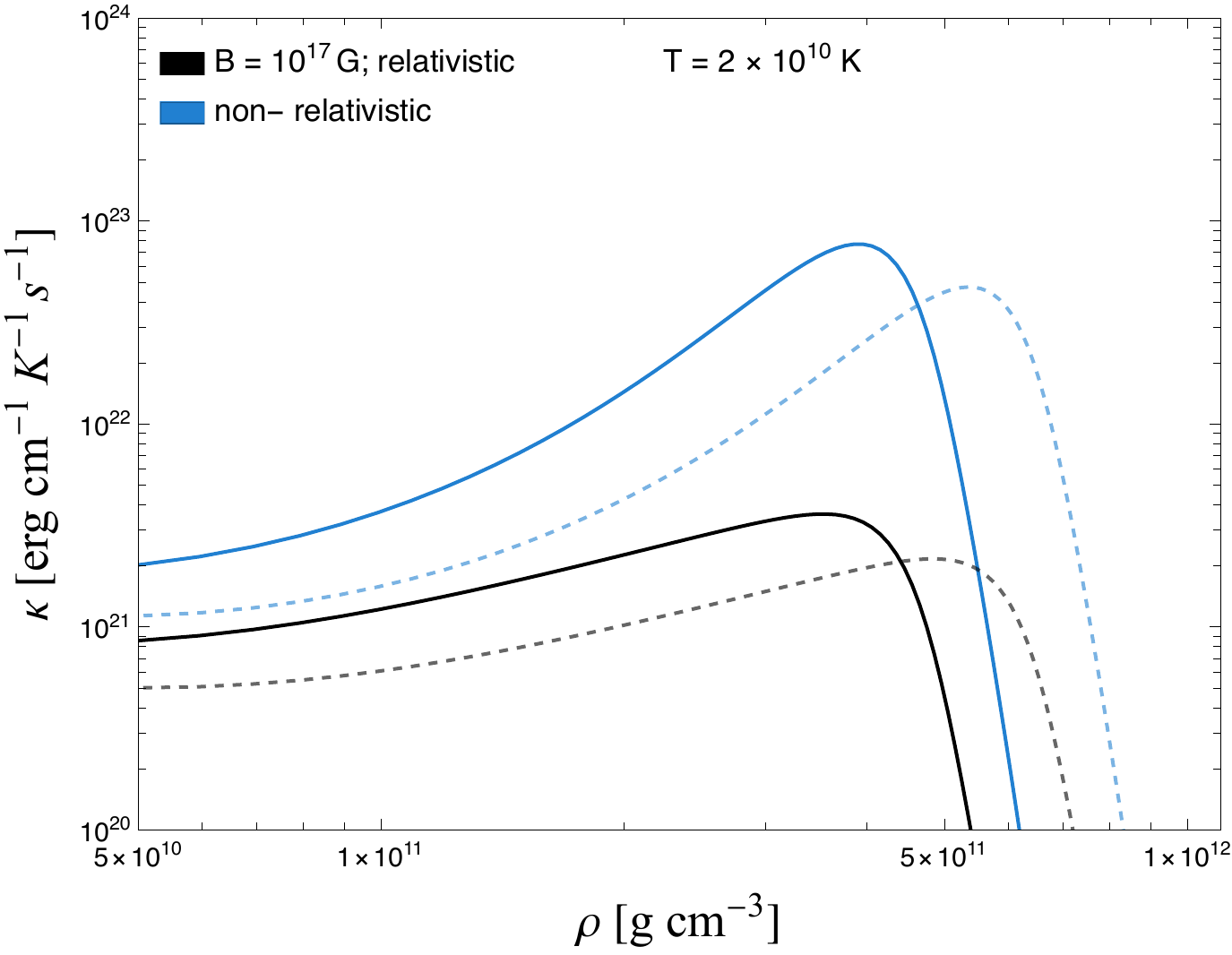}~~~~~\includegraphics[width=0.4\textwidth]{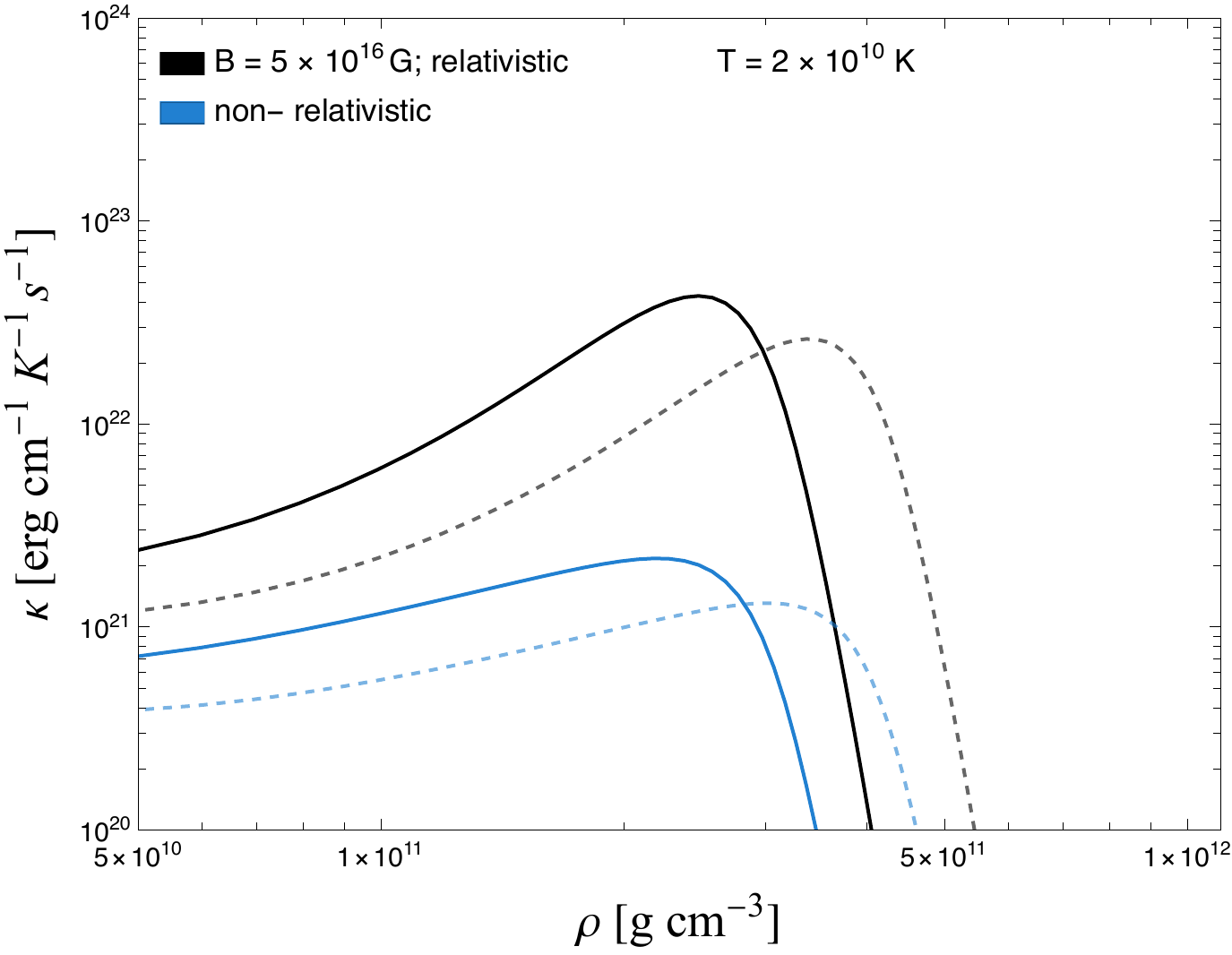}
\subcaption{Effect of HDL propagator on $\kappa$}
\end{subfigure}
\caption{Plots  of  $\sigma$ (upper panel) and $\kappa$ (lower panel) with $\rho$ comparing static, Coulomb screening in non-relativistic plasma (blue) and dynamical screening in relativistic plasma (black)  at  fixed temperature of $2 \times 10^{10} $ K. The left panel curves  are for B$=10^{17}$ G and the right curves are for B$=5\times 10^{16}$ G.  The choice of elements are Mo  (dotted lines) and Fe (solid lines).}
\label{fig:sigmakapparelnonrel}
\end{figure*}
\vskip 12pt
\begin{figure*}
\centering
\includegraphics[width=0.4\textwidth]{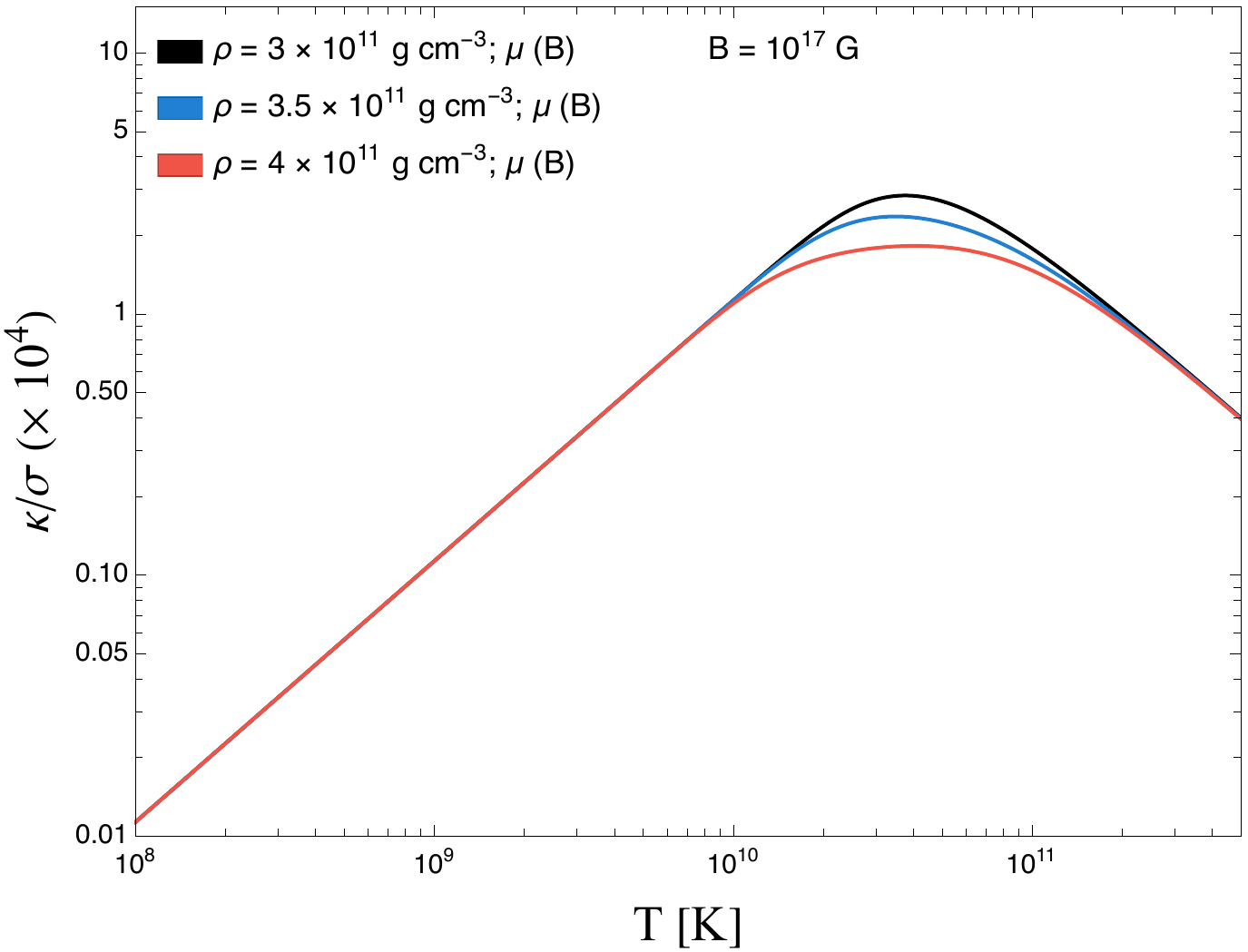}
\caption{The variation of $\kappa/\sigma$ with $T$ for different values of densities $3\times 10^{11} $ gcm$^{-3}$, $3.5\times 10^{11}$  gcm$^{-3}$, $4\times 10^{11}$  gcm$^{-3}$. The chosen magnetic field  is $10^{17}$ G. }
\label{fig:ratiosigmabykappa}
\end{figure*}
 \subsection{Effect of HDL modified propagator}
 In the Fig.\ref{fig:sigmakapparelnonrel} we explicitly estimate the modification induced in $\sigma$ and $\kappa$ when electron-ion interact through HDL photon propagator and also compare the case when these plasma constituents interact through static, longitudinal photon propagator. The upper panel is for variation of $\sigma$ and the lower panel is for $\kappa$. The left panel considers magnetic field of B=$10^{17}$ G and the right panel B=$5\times 10^{16}$ G. In all the four plots we consider $T=2\times 10^{10}$K. Blue curves represent the coefficients when the interaction rate is screened through static, longitudinal photon propagator relevant in non-relativistic plasma and black curves when the constituents interact through frequency dependent HDL photon propagator required in the relativistic plasma.   We find that the inclusion of HDL propagator  reduces the values of both the coefficients  substantially. For $\sigma$ the maximum value of the curve decreases by a factor of $31$ in comparison to the case when the interaction rate is Debye screened, whereas, for $\kappa$ the factor of reduction is $20$. 

It is known that modifications to different equilibrium and non-equilibrium properties of plasma due to inclusion of HDL propagators emerge from frequency dependent photon propagator. 
The reduction in both the transport coefficients shown above arises due to the increase in the interaction rate caused by non-zero  frequency in both the longitudinal and transverse components of the photon propagator. The electromagnetic perturbation to the distribution function ($\Psi$) is inversely proportional to $ I_{fi}$ and $\sigma, \kappa$ are directly proportional to $\Psi$. Hence, increment in $\Psi$ decreases both the coefficients. The increment in $I_{fi}$ with the inclusion of HDL photon propagator can be seen from Eq.(\ref{eq:phi}). The second term in  Eq.(\ref{eq:phi})  is dependent on $v_k<1$. $v_k$   reduces  the numerator of second term in  Eq.(\ref{eq:phi}). In  the denominator of the same equation  $q_{\perp}^2<2m\omega_B$  and $m_D^2<2m\omega_B$ lead to  $u<1$, $\zeta<1$ respectively. Hence, the reduction due to small value of $v_k$ gets compensated and interaction rate increases due to frequency dependent dynamical screening. 
\subsection{Estimation of dissipation time scales}
In this paper we give quantitative estimations of  both the time scales related to electrical and thermal transport coefficients.  The magnetic field decay (or diffusion) timescale due to Ohmic dissipation is given by the well-known expression \cite{Harutyunyan:2018mpe, Rezzolla_book:2013},
$\tau_{\sigma} = 4\pi\sigma \lambda^2_B/c^2$.  The time scale is dependent on the two factors magnetic field scale height  ($\lambda_B$) and $\sigma$. For a typical value of $\lambda_B\sim 10^{-4}$ c.m (  $\lambda_B\geq\lambda_{mfp}\mbox{(electron mean free path)}\sim 10^{-4}$cm, $\lambda_B< \lambda_{lim} \mbox{(limiting magnetic field scale height)}\sim 10^{-3}$cm)   we obtain $\tau_{\sigma}\sim 15$ $ms$ which is well within the range of survival time period of neutron star merger. Choice of $\lambda_B$ satisfies both the conditions specified by Ref.\cite{Harutyunyan:2018mpe}, $\lambda_{mfp}<\lambda_B>\lambda_{lim}$. With the same choice of $\lambda_B$,  $\tau_{\sigma}\sim 600$ $ms$ in the non-relativistic scenario. We emphasize on the reduction of time scale due to inclusion of dynamical screening in the calculation in contrast to the scenario when electrons-ions interact through longitudinal plasmon. 

Thermal conduction time scale is defined as $\tau_{\kappa}=c_v \Delta z^2/6\kappa$, where, $c_v$ is the specific heat and $\Delta z$ is the region which is hotter than surrounding by a temperature difference $\Delta T$. The  contribution of degenerate relativistic electrons in specific heat is, 
$c_v=5.4\times 10^{19}\left(\frac{n_e}{n_0}\right)^{2/3}T_9$,  $n_0$ is the nuclear saturation density and $T_9=T/10^9$. For $n_e=0.57\times 10^{35} \mbox{cm}^{-3}$, temperature $T=2\times 10^{10}$K and $\Delta z\sim 25$ c.m, thermal equilibration time scale becomes $\sim 50$ ms which is of the order of time scale of the merged object. For $\Delta z\sim 1$ km, the time scale is much greater than the survival time period of the merged compact star.  However, the  estimation of time scales presented here demands a more detailed hydrodynamic  calculation  to assess the inclusion of   dissipation  and resistivity in magneto-hydrodynamic simulation of BNS mergers.

\subsection{Validity of Wiedemann-Franz law}
 In a typical degenerate plasma, electrons' conduction dominates system's dissipation coefficients through electrical and thermal coefficients. This suggests linear increment in $\sigma/\kappa$ with $T$ known as Wiedemann-Franz law. From the plot of $\sigma$ in fig.(\ref{fig:sigmakappavsTcomb}), we see that when $T<5\times 10^{11}$ K (for $\rho=4\times 10^{11}$gm cm$^{-3}$) $\sigma$ is temperature independent. For  $T>5\times 10^{11}$ K, $\sigma$ decreases with temperature. However, $\kappa$ increases with temperature and after the mentioned temperature, it decreases. Hence, $\kappa/\sigma$ first increases with temperature and after $5\times 10^{11}$ K it decreases, violating Wiedemann-Franz law. From eq.(\ref{sigma_final}) and eq.(\ref{kappa_final}) it is evident that if $(\epsilon_p-\mu)= T$  linear scaling of temperature exists between $\kappa$ and $\sigma$. At temperatures less than $T<5\times 10^{10}$ K, $(\epsilon_p-\mu)< T$, hence, $\sigma/\kappa\propto T^3$. In both the temperature domains of $T<5\times 10^{10}$K and $T>5\times 10^{10}$K, degeneracy condition $|\epsilon_p-\mu|\ll \mu$ gets satisfied. 
\section{Summary and conclusions}
In this work, we have calculated the quantized longitudinal electronic transport coefficients  in dynamically screened hot and dense magnetized QED plasma involved in binary neutron star merger simulations.
   The calculation considers  scattering of electrons with ions through screened electromagnetic force in electron-ion plasma. We have presented the plots for the variation of transport coefficients with density ( $\sim 10^{12}$ gm cm$^{-3}$), temperature ($\sim 10^{10}$ K) and magnetic field ($\sim 10^{17}$ G) for two elements Mo and Fe. The scales for the generation of the plots are chosen so that they obey the conditions of relativistic electrons at density $\rho>10^{6}$ g cm$^{-3}$, temperature $T> 5.93\times 10^{9}$ K and zeroth Landau level population by obeying $\sqrt{E^2-1}/2b\ll 1$. These two  constraints indicate, high magnetic field and low density regime of BNS mergers as the valid physical domain of our calculation.

For the calculation of both electrical and thermal conductivities, we have assumed particles are slightly out of equilibrium which allows us to solve the Boltzmann equation numerically. We calculate electron-ion scattering amplitude with screened electromagnetic interaction of magnetically modified spinors. The off-equilibrium distribution function has been obtained by solving the Boltzmann kinetic equation in relaxation time approximation. However, we have not considered the finite size of the nuclei and ion structure function for the calculation of the relaxation rate.

The electromagnetic interaction between electrons and ions have been incorporated through HDL propagator in the calculation. The calculation should account for magnetically modified anisotropic  photon propagator; however in the present paper, we have considered only isotropic HDL propagator. For the screening mass zero temperature  magnetically modified  Debye mass has been incorporated  but in the hot and dense plasma, inclusion of finite temperature screening mass would be more relevant.

We have found that the inclusion of the HDL propagator in the relativistic plasma reduces the values of both $\sigma$ and $\kappa$ in contrast to static screening in the non-relativistic plasma.  The frequency dependent screening in the HDL propagator enhances the interaction rate. 
 The off equilibrium distribution function is inversely proportional to the interaction rate and the transport coefficients are directly proportional to $\Phi$. Hence,  enhancement in the interaction rate decrease  both the transport coefficients  at high density in the degenerate regime. We also observe cubic temperature scaling between $\kappa$ and $\sigma$ violating the Wiedemann-Franz law. 

A rough estimation of  diffusion and thermal equilibration time scales from the expressions of quantized $\sigma$, $\kappa$ including frequency dependent screening have also been attempted.  The obtained Ohmic time scale has  found out to be of the same order as the survival time of the merged compact object due to reduction in the value of $\sigma$ (for $\lambda_B\ll$ cm). The thermal equilibration time scale can also be found to match with the time scale of the merged object when $\Delta z \sim $ cm.  One can infer from the analysis that many-body effects play an important role in determining the dissipative time scales relevant in the neutron star merger in the low density, high temperature and high magnetic field  regime if the length-scales are of the order mentioned above. This  estimation of dissipation time scales of the transport coefficients requires rigorous hydrodynamic formulation for more realistic values of $\lambda_B$ and $\Delta z$. The current calculation of $\sigma$ and $\kappa$ can be implemented in modeling the magneto-thermal evolution of the merged compact object as well.

In the current paper, we have considered only longitudinal component of the transport coefficients (and ignored all the other tensorial components) in the equation for magnetic-field evolution in plasma. The realistic estimate of $\tau_{\sigma}$ and $\tau_{\kappa}$ can only be obtained if all the components of conductivity tensor are known in the background of relevant equation of state for neutron star merger. However, our calculations presented in this paper provide a significant step towards conjoining the complex microphysical effects in plasma with the GRMHD simulations. The present formalism of transport theory can be extended in  relativistic, magnetised QCD  matter with certain modifications like inclusion of QCD coupling constant,  diagrams of strong interaction 
and proper vertex corrections. 
\subsection*{Acknowledgments}
 Authors would also like to thank and acknowledge R. Nandi for  fruitful discussions regarding various aspects of this work. Authors acknowledge fruitful discussion with T. Mazumder regarding  numerical analysis of the current project. SPA acknowledges the Polish Academy of Sciences through grant agreement \\ PAN.BFD.S.BDN.612.022.2021-PASIFIC 1, QGPAnatomy. This work received funding from the European Union’s Horizon 2020 research and innovation program under the Maria Sklodowska-Curie grant agreement No.\ 847639 and from the Polish Ministry of Education and Science.
\section*{Appendix A}
\label{app:A}
In this appendix, we present the important steps for the evaluation of electron-ion scattering rate in presence of magnetic field. An electron of momentum $p\equiv(\epsilon_p, \vec {p})$ scatters with an ion of momentum $k\equiv(\epsilon_k, \vec {k})$ leading to the final momentum states $p'\equiv(\epsilon_{p'}, \vec {p'})$ and $k'\equiv(\epsilon_{k'}, \vec {k'})$. The scattering rate from initial state  to final state in absence of magnetic field is given by,
\bea\label{eq:Ifi1}
&&I_{fi}=\frac{1}{2\epsilon_p}\int\frac{d^3p'}{(2\pi)^32\epsilon_{p'}}\int\frac{d^3k}{(2\pi)^32\epsilon_k}\int\frac{d^3k'}{(2\pi)^32\epsilon_{k'}}\nn\\
&&[f_{0}(\epsilon_p)g_{0}(\epsilon_k)\l(1-f_{0}(\epsilon_p')\r)-f_{0}(\epsilon_p')g_{0}(\epsilon_k')\l(1-f_{0}(\epsilon_p)\r) ]\nn\\
&&\l(2\pi \r)^4\delta\l(\epsilon_p+\epsilon_k-\epsilon_{p'}-\epsilon_{k'}\r)\delta^3\l(\vec {p}+\vec {k} -\vec {p'}-\vec {k'}\r)|{\cal M}_{fi}|^2\nn\\
 \eea
The above equation is modified in presence of  magnetic field,
\bea\label{eq:Ifi3}
I_{fi}&=&\frac{eB}{16(2\pi)^5}\sum_{n',p'_z,s'}\int dp'_zdy_B\int d^3kdq_y \int\frac{d\Omega_k}{4\pi}\nn\\
&&g_f(k)\l(\Phi_{n,p_z,s}-\Phi_{n',p'_z, s'}\r)\delta(\epsilon_p+\epsilon_k-\epsilon_p'-\epsilon_k')\nn\\
&&\sum_{q_x,q_z}\delta_{k_x'-k_x,q_x}\delta_{k_z'-k_z,q_z}|{\cal M}_{fi}|^2,
\label{int_rate2}
\eea
where, $y_B=p_x/m\omega_B$ and $\Phi_{n,p_z,s}$ has already been defined in the Introduction. We have inserted $\int d\Omega_k/4\pi=1$ in the above equation.
The argument of delta function can be written as follows,
\bea\label{eq:deltaeps}
\delta(\epsilon_p+\epsilon_k-\epsilon_{p'}-\epsilon_{k'})&=&\delta(\epsilon_{k}-\epsilon_{k'}-\hat p.\vec{q})
\eea
where, we have used $\epsilon_{|p'|}=\epsilon_{|p-q|}=\epsilon_p-\hat p.\vec{q}$. The angular integrals of $\vec{k}$ can be expressed as,
\bea\label{eq:angint1}
&&\int \frac{d\Omega_k}{4\pi}\delta(\hat p.\vec{ q}-\vec{v_k}.\vec{ q})=\frac{1}{2|\vec {q}|}\nn\\
&&\int \frac{d\Omega_k}{4\pi}\delta(\hat p.\vec{ q}-\vec{v_k}.\vec{ q})(\hat p.\hat k-\hat p.\hat q \hat q.\hat k)^2\nn\\
&&=\frac{1}{4|\vec{q}|}\left(1-\frac{(\hat p.\vec{ q})^2}{\vec {|q|}^2}\right)\left(v_k^2-\frac{(\hat p.\vec{ q})^2}{\vec {|q|}^2}\right).
\eea
 Eq.(\ref{int_rate2}) can further be written as,
\bea\label{eq:Ifi4}
I_{fi}&=&\frac{n_i}{32(2\pi)^2 }\sum_{n',p_z',s'}\int dq_zdq_x\frac{ dq_y }{\vec{|q|}}\nn\\
&&\l[\Phi_{n,p_z,s}-\Phi_{n',p'_z, s'}\r] |{\cal M}_{fi}|^2,
\eea
where, we have changed the variable $dy_B'= dq_x/m\omega_B$ following momentum conservation $p'_x-p_{x}=q_x$. In the above equation, $n_i$ is the number density of ions  which can  be expressed in terms of electron number density as $n_i=n_e/Z$. Further, $n_e$  can be expressed in terms of the Debye mass as $n_e=\mu m_D^2/3 e^2$.

Next we introduce a dimensionless variable  $y=q_z/|\vec{q}|$,
\bea\label{eq:Ifi4}
I_{fi}&=&\frac{\mu m_D^2 eB}{3\times 32 Z e^2 (2\pi)^2 }\sum_{n',p_z',s'}\int dydq_x dq_y\nn\\
&&[\Phi_{n,p_z,s}-\Phi_{n',p'_z, s'}]|{\cal M}_{fi}|^2.
\label{int_rate}
\eea
In order to calculate $|{\cal M}|^2$, we use following electronic spinor in presence of magnetic field, 
\begin{equation}\label{eq:up}
\Psi(r) =\frac{exp[i(p_xx+p_zz)]}{\sqrt{L_x L_z}}\Biggl(\begin{matrix} 
\tilde\alpha \tilde AH_{n-1}(\xi) \\-s\tilde\alpha \tilde \beta \tilde H_n(\xi) \\s\tilde\beta \tilde A \tilde H_{n-1}(\xi) \\\tilde \beta \tilde B \tilde H_{n}(\xi)
 \end{matrix}
 \Biggr).
\end{equation}
Using above spinors and  the expression for photon propagator (Eq.(\ref{PiT})) in Eq.(\ref{mat_amp}) in Section.\ref{formalism}  one obtains,
\bea\label{intmatrix}
&&\sum_{spin} |{\cal M}|^2=\l(4\pi Ze^2\r)^2\bigg[\frac{2\pi ym_D^2}{\l(q_{\perp}^2+Re \Pi_L\r)^2+Im \Pi_L^2}\nn\\
&&-\frac{2\pi y m_D^2v_k^2}{2(q_{\perp}^2+Re \Pi_T)^2+Im \Pi_T^2)}\bigg]\sum_{n',p_z',s'}\l[ss'\tilde \alpha^2+\tilde \beta^2\r]\nn\\
&&\l[ss'\tilde A \tilde A'I_{n'-1}I_{n-1}+\tilde A \tilde A'I_{n'-1}I_{n-1}\r]^2
\label{mat_amp2},
\eea 
where, $s$ and $s'$ are $\pm$, 
\begin{equation}\label{eq:alphabeta}
 \Bigg(\begin{matrix} 
\tilde \alpha \\ \tilde \beta\\
 \end{matrix}
 \Bigg)=\Bigg(\begin{matrix} 
\sqrt{\frac{1}{2}(1+\frac{m}{\epsilon_p})}\\ \sqrt{\frac{1}{2}(1-\frac{m}{\epsilon_p})}\\
 \end{matrix}
 \Bigg),
\end{equation}
\begin{equation}\label{eq:ABmatrix}
 \Bigg(\begin{matrix} 
\tilde A\\ \tilde  B\\
 \end{matrix}
 \Bigg)=\Bigg(\begin{matrix} 
 [\frac{1}{2}(1+\frac{sp_z}{\sqrt{\epsilon_p^2-m^2}})]^{1/2}\\
[\frac{1}{2}(1-\frac{-sp_z}{\sqrt{\epsilon_p^2-m^2}})]^{1/2}\\ 
 \end{matrix}
 \Bigg)
\end{equation}
and
\bea\label{eq:hermitepol1}
I_{n',n}=\int_{-\infty}^{\infty} exp(iq_y y)\tilde H_{n'}(\xi')\tilde H_n(\xi)dy.
\eea
$H_n(\xi)$ is the Hermite polynomial,
\bea\label{eq:hermitepol2}
H_n(\xi)=\frac{m\omega_B}{\pi }^{\frac{1}{4}}(2^n n!)^{\frac{-1}{2}}exp^{\frac{-\xi^2}{2}}H_n(\xi),
\eea

$\xi=\sqrt{m\omega_B}$.
Inserting $I_{n',n}$  in Eq.(\ref{mat_amp2}) and performing the $y$ integration we obtain,

\bea
&&\sum_{spin} |{\cal M}|^2 =\l(4\pi Ze^2\r)^2\Big[\frac{2\pi ym_D^2}{(q_{\perp}^2+Re \Pi_L)^2+Im \Pi_L^2}\nn\\
&-&\frac{2\pi y m_D^2v_k^2}{2(q_{\perp}^2+Re \Pi_T)^2+Im \Pi_T^2)}\Big]\sum_{n',p_z',s'}\Big[1+ss'\nn\\
&&+\frac{m^2}{\epsilon_p^2}(1-ss')\Big[1+ss'-
\frac{1}{2}\frac{ss'}{\epsilon_p^2-m^2}(\eta'p_{n'}-p_z)^2\Big]\nn\\
&&\Big[F^2_{n',n}(u)+F^2_{n'-1, n-1}(u)]\Big]-
\frac{ss'u \omega_B m}{\epsilon_p^2-m^2}\nn\\
&&\Big[F^2_{n'-1, n}(u)+F^2_{n', n-1}(u)\Big]\nn\\
&&-\frac{sp_z+s'\eta'p_n'}{\sqrt{\epsilon_p^2-m^2}}\Big[F^2_{n', n}(u)-F^2_{n'-1, n-1}(u)\Big]
\label{mat_amp3}
\eea
where, $\eta'=\pm$, $F_{n',n}(u)=exp^{-u/2}u^{\frac{n-n'}{2}}\sqrt{\frac{n'!}{n!}}L_{n'}^{n-n'}$ and $u=\frac{1}{2m\omega_B}(q_x^2+q_y^2)$. The functions  $L_{n'}^{n-n'}(u)$ are Laguerre polynomials and  $F_{n',n}(u)$ are normalized to \\$\int_0^{\infty}F_{n',n}^2 du=1$.


Now, to perform the  integration in $y$ in Eq.(\ref{int_rate}) we use the following sum rule,
\bea\label{eq:sumrule}
  &&\int_{-1}^{1} \frac{dy}{y}\, \frac{1}{2\pi}\, 
  \frac{2\,{\rm Im}\,\Pi_{L}(y)}
       {(q_{\perp}^2{+}{\rm Re}\,\Pi_{L}(y))^2 + ({\rm Im}\,\Pi_{L}(y))^2}\nn\\
       &&-\frac{2v_k^2\,{\rm Im}\,\Pi_{T}(y)}
       {(q_{\perp}^2{+}{\rm Re}\,\Pi_{T}(y))^2 + ({\rm Im}\,\Pi_{T}(y))^2}\nn\\
&& =\left( \frac{1}{q_{\perp}^2+{\rm Re}\,\Pi_{T,L}(y{=}\infty)}
        -\frac{v_k^2}{q_{\perp}^2+{\rm Re}\,\Pi_{T,L}(y{=}0)}\right). 
        \nn\\
        \eea
      
 In the limiting case, ${\rm Re}\,\Pi_{T,L}(y{=}\infty) = m_D^2/3$,\\
 ${\rm Re}\,\Pi_{T}(y{=}0)=0, 
 {\rm Re}\,\Pi_{L}(y{=}0)=m_D^2$.

Using the above relations the interaction rate becomes,
\bea\label{Ififinal}
&&I_{fi}=\frac{n_i}{32(2\pi)^2 }\int  dq_xdq_y \l(\Phi_{n,p_z,s}-\Phi_{n',p'_z, s'}\r)\nn\\
&&\left[\frac{2}{3(q_{\perp}^2+\frac{m_D^2}{3})(q_{\perp}^2+m_D^2)}-\frac{v_k^2}{6q_{\perp}^2(q_{\perp}^2+\frac{m_D^2}{3})}\right]
{\cal F},
\eea
where,
\bea\label{eq:F}
&&{\cal F}=\frac{4\pi \sigma_0}{m^2}\sum_{n',p_z',s'}[1+ss'
+\frac{m^2}{\epsilon_p^2}(1-ss')\nn\\
&&[1+ss'-
\frac{1}{2}\frac{ss'}{\epsilon_p^2-m^2}(\eta'p_{n'}-p_z)^2]\nn\\
&&[F^2_{n',n}(u)+F^2_{n'-1, n-1}(u)]]\nn\\
&&-\frac{ss'u \omega_B m}{\epsilon_p^2-m^2}
[F^2_{n'-1, n}(u)+F^2_{n', n-1}(u)]\nn\\
&&-\frac{sp_z+s'\eta'p_n'}{\sqrt{\epsilon_p^2-m^2}}[F^2_{n', n}(u)-F^2_{n'-1, n-1}(u)].
\eea
We change the variable $q_y$ to $u$ and perform the integration as follows,
\bea\label{eq:qintegral}
\int dq_xdq_y&=& m\omega_B \int \frac{dq_x du}{\sqrt{2m\omega_Bu-q_x^2}}\nn\\
&=&m\omega_B \pi\int du
\eea
Finally, the particle scattering rate becomes,
\bea\label{eq:Ififinal2}
I_{fi}&=&\frac{n_i }{2}\sum_{n',p_z',s'}\int  du\l(\Phi_{n,p_z,s}-\Phi_{n',p'_z, s'}\r)\nn\\
&&\l[\frac{1 }{3(u+\frac{\zeta}{3})(u+\zeta)}-\frac{v_k^2}{6u(u+\frac{\zeta}{3})}\r]{\cal F},
\end{eqnarray}
where, $\zeta=m_D^2/2m \omega_B$.
\section*{Appendix B}
\label{app:B}
\begin{figure*}[h]
\centering
\includegraphics[width=0.4\textwidth]{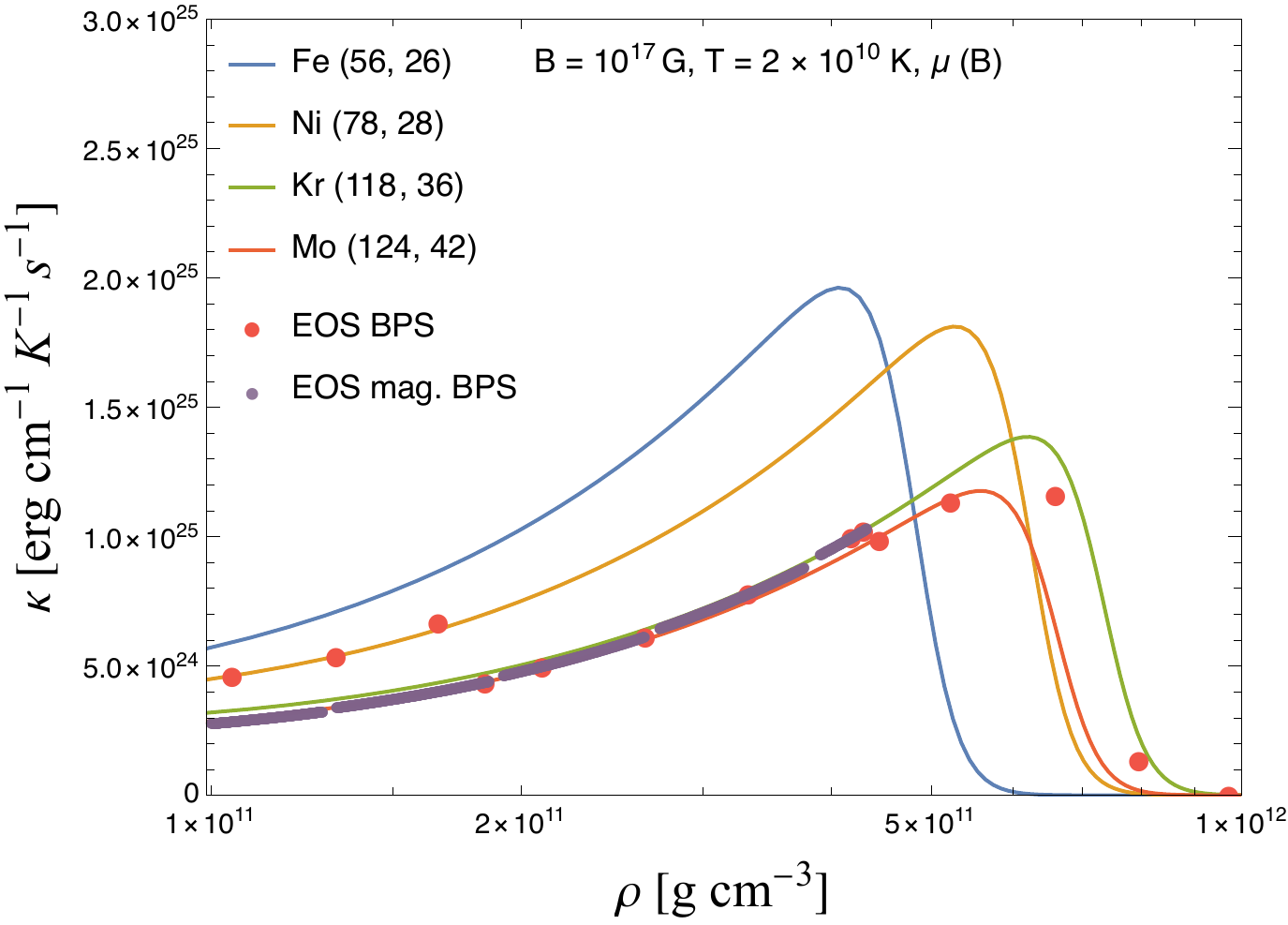}~~~~~\includegraphics[width=0.4\textwidth]{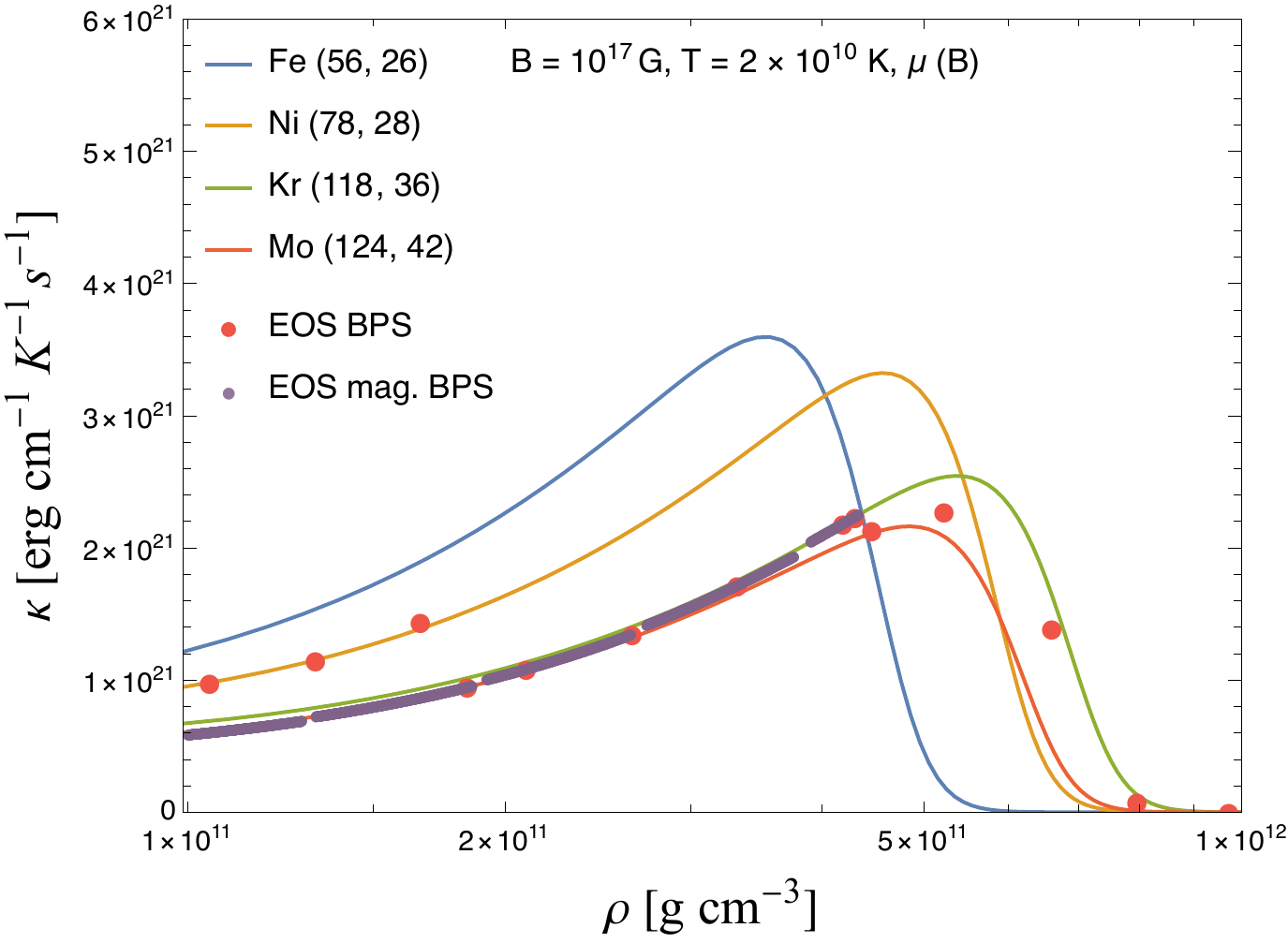}
\caption{The variation of $\sigma$ (left) and $\kappa$ (right) with $\rho$  for different elements. The magnetic field is chosen as $10^{17} $G and temperature as $2 \times 10^{10} $ T. We  show the behaviour of the transport coefficients  for two different  EOS s given in Refs. {\cite{Baym:1971pw, Nandi:2010fp}}}
\label{eos_comp}
\end{figure*}

In this appendix, we  show the compariosn of the conductivities considering two equation of states (EOSs), BPS (Baym, Pethick and Surtherland) \cite{Baym:1971pw} and magnetic BPS model \cite{Nandi:2010fp} and comment on considering  Fe and Mo for numerical analysis in the paper. In ref.\cite{Baym:1971pw} the equation of state of zero-temperature matter in complete nuclear equilibrium is given for mass densities below $5\times 10^{14}$ g cm$^{-3}$.  In Ref.\cite{Nandi:2010fp}, the BPS \cite{Baym:1971pw} has been extended to include the physical parameters for a low density plasma in presence of high magnetic field relevant for neutron star crust. 
 In the fig.\ref{eos_comp}, we  plot $\sigma$ and $\kappa$ with $\rho$ for these two different EOSs. From the plot fig.\ref{eos_comp} it is evident that the BPS curve matches well with the Mo curve for both $\sigma$ and $\kappa$. We also include Fe for reference.



\begin{thebibliography}{1}
\bibitem{Dexheimer:2020zzs}
V.~Dexheimer, J.~Noronha, J.~Noronha-Hostler, C.~Ratti and N.~Yunes,
J. Phys. G \textbf{48} (2021) no.7, 073001
doi:10.1088/1361-6471/abe104
[arXiv:2010.08834 [nucl-th]].
\bibitem{LIGOScientific:2017vwq}
B.~P.~Abbott \textit{et al.} [LIGO Scientific and Virgo],
Phys. Rev. Lett. \textbf{119} (2017) no.16, 161101
doi:10.1103/PhysRevLett.119.161101
[arXiv:1710.05832 [gr-qc]].

\bibitem{LIGOScientific:2017zic}
B.~P.~Abbott \textit{et al.} [LIGO Scientific, Virgo, Fermi-GBM and INTEGRAL],
Astrophys. J. Lett. \textbf{848} (2017) no.2, L13
doi:10.3847/2041-8213/aa920c
[arXiv:1710.05834 [astro-ph.HE]].

\bibitem{LIGOScientific:2017ync}
B.~P.~Abbott \textit{et al.} [LIGO Scientific, Virgo, Fermi GBM, INTEGRAL, IceCube, AstroSat Cadmium Zinc Telluride Imager Team, IPN, Insight-Hxmt, ANTARES, Swift, AGILE Team, 1M2H Team, Dark Energy Camera GW-EM, DES, DLT40, GRAWITA, Fermi-LAT, ATCA, ASKAP, Las Cumbres Observatory Group, OzGrav, DWF (Deeper Wider Faster Program), AST3, CAASTRO, VINROUGE, MASTER, J-GEM, GROWTH, JAGWAR, CaltechNRAO, TTU-NRAO, NuSTAR, Pan-STARRS, MAXI Team, TZAC Consortium, KU, Nordic Optical Telescope, ePESSTO, GROND, Texas Tech University, SALT Group, TOROS, BOOTES, MWA, CALET, IKI-GW Follow-up, H.E.S.S., LOFAR, LWA, HAWC, Pierre Auger, ALMA, Euro VLBI Team, Pi of Sky, Chandra Team at McGill University, DFN, ATLAS Telescopes, High Time Resolution Universe Survey, RIMAS, RATIR and SKA South Africa/MeerKAT],
Astrophys. J. Lett. \textbf{848} (2017) no.2, L12
doi:10.3847/2041-8213/aa91c9
[arXiv:1710.05833 [astro-ph.HE]].




\bibitem{Paschalidis:2016agf}
V.~Paschalidis,
Class. Quant. Grav. \textbf{34} (2017) no.8, 084002
doi:10.1088/1361-6382/aa61ce
[arXiv:1611.01519 [astro-ph.HE]].
\bibitem{Kawamura:2016nmk}
T.~Kawamura, B.~Giacomazzo, W.~Kastaun, R.~Ciolfi, A.~Endrizzi, L.~Baiotti and R.~Perna,
Phys. Rev. D \textbf{94} (2016) no.6, 064012
doi:10.1103/PhysRevD.94.064012
[arXiv:1607.01791 [astro-ph.HE]].
\bibitem{Ruiz:2016rai}
M.~Ruiz, R.~N.~Lang, V.~Paschalidis and S.~L.~Shapiro,
Astrophys. J. Lett. \textbf{824} (2016) no.1, L6
doi:10.3847/2041-8205/824/1/L6
[arXiv:1604.02455 [astro-ph.HE]].

\bibitem{Palenzuela:2013hu}
C.~Palenzuela, L.~Lehner, M.~Ponce, S.~L.~Liebling, M.~Anderson, D.~Neilsen and P.~Motl,
Phys. Rev. Lett. \textbf{111} (2013) no.6, 061105
doi:10.1103/PhysRevLett.111.061105
[arXiv:1301.7074 [gr-qc]].

\bibitem{Baiotti:2016qnr}
L.~Baiotti and L.~Rezzolla,
Rept. Prog. Phys. \textbf{80} (2017) no.9, 096901
doi:10.1088/1361-6633/aa67bb
[arXiv:1607.03540 [gr-qc]].

\bibitem{Anderson:2008zp}
M.~Anderson, E.~W.~Hirschmann, L.~Lehner, S.~L.~Liebling, P.~M.~Motl, D.~Neilsen, C.~Palenzuela and J.~E.~Tohline,
Phys. Rev. Lett. \textbf{100} (2008), 191101
doi:10.1103/PhysRevLett.100.191101
[arXiv:0801.4387 [gr-qc]].

\bibitem{Liu:2008xy}
Y.~T.~Liu, S.~L.~Shapiro, Z.~B.~Etienne and K.~Taniguchi,
Phys. Rev. D \textbf{78} (2008), 024012
doi:10.1103/PhysRevD.78.024012
[arXiv:0803.4193 [astro-ph]].

\bibitem{Palenzuela:2008sf}
C.~Palenzuela, L.~Lehner, O.~Reula and L.~Rezzolla,
Mon. Not. Roy. Astron. Soc. \textbf{394} (2009), 1727-1740
doi:10.1111/j.1365-2966.2009.14454.x
[arXiv:0810.1838 [astro-ph]].

\bibitem{Dionysopoulou:2012zv}
K.~Dionysopoulou, D.~Alic, C.~Palenzuela, L.~Rezzolla and B.~Giacomazzo,
Phys. Rev. D \textbf{88} (2013), 044020
doi:10.1103/PhysRevD.88.044020
[arXiv:1208.3487 [gr-qc]].

 
\bibitem{Dionysopoulou:2015tda}
K.~Dionysopoulou, D.~Alic and L.~Rezzolla,
Phys. Rev. D \textbf{92} (2015) no.8, 084064
doi:10.1103/PhysRevD.92.084064
[arXiv:1502.02021 [gr-qc]].


\bibitem{Kiuchi:2015qua}
K.~Kiuchi, Y.~Sekiguchi, K.~Kyutoku, M.~Shibata, K.~Taniguchi and T.~Wada,
Phys. Rev. D \textbf{92} (2015) no.6, 064034
doi:10.1103/PhysRevD.92.064034
[arXiv:1506.06811 [astro-ph.HE]].

\bibitem{Kiuchi:2017zzg}
K.~Kiuchi, K.~Kyutoku, Y.~Sekiguchi and M.~Shibata,
Phys. Rev. D \textbf{97} (2018) no.12, 124039
doi:10.1103/PhysRevD.97.124039
[arXiv:1710.01311 [astro-ph.HE]].
  
 

 
 
\bibitem{Ruiz:2017due}
M.~Ruiz, S.~L.~Shapiro and A.~Tsokaros,
Phys. Rev. D \textbf{97} (2018) no.2, 021501
doi:10.1103/PhysRevD.97.021501
[arXiv:1711.00473 [astro-ph.HE]].


\bibitem{Palenzuela:2013kra}
C.~Palenzuela, L.~Lehner, S.~L.~Liebling, M.~Ponce, M.~Anderson, D.~Neilsen and P.~Motl,
Phys. Rev. D \textbf{88} (2013) no.4, 043011
doi:10.1103/PhysRevD.88.043011
[arXiv:1307.7372 [gr-qc]].

\bibitem{Harutyunyan:2018mpe}
A.~Harutyunyan, A.~Nathanail, L.~Rezzolla and A.~Sedrakian,
Eur. Phys. J. A \textbf{54} (2018) no.11, 191
doi:10.1140/epja/i2018-12624-1
[arXiv:1803.09215 [astro-ph.HE]].

\bibitem{Harutyunyan:2016rxm}
A.~Harutyunyan and A.~Sedrakian,
Phys. Rev. C \textbf{94} (2016) no.2, 025805
doi:10.1103/PhysRevC.94.025805
[arXiv:1605.07612 [astro-ph.HE]].

\bibitem{Alford:2017rxf}
M.~G.~Alford, L.~Bovard, M.~Hanauske, L.~Rezzolla and K.~Schwenzer,
Phys. Rev. Lett. \textbf{120} (2018) no.4, 041101
doi:10.1103/PhysRevLett.120.041101
[arXiv:1707.09475 [gr-qc]].


\bibitem{Aguilera:2007xk}
D.~N.~Aguilera, J.~A.~Pons and J.~A.~Miralles,
Astron. Astrophys. \textbf{486} (2008), 255-271
doi:10.1051/0004-6361:20078786
[arXiv:0710.0854 [astro-ph]].

\bibitem{Yakovlev1980} D.~G. Yakovlev and  V.~A. Urpin, 
Sov. Astron. {\bf 3} (1980) 24.

 \bibitem{Hernquist1984} L. Hernquist, 
 Astrophys. J. {\bf 56} (1984) 325.

\bibitem{Potekhin:1996zh}
A.~Y.~Potekhin,
Astron. Astrophys. \textbf{306} (1996), 999-1010
[arXiv:astro-ph/9603133 [astro-ph]].
 
\bibitem{Potekhin:1996hu}
A.~Y.~Potekhin and D.~G.~Yakovlev,
Astron. Astrophys. \textbf{314} (1996), 341
[arXiv:astro-ph/9604130 [astro-ph]].

\bibitem{1964Abrikosov}
A.~A. {Abrikosov}, 
 Soviet
  Physics JETP {\bf 18} (1964) 1399--1404. 
  
\bibitem{1966ApJ...146..858H}
W.~B. {Hubbard}, 
 Astrophys. J {\bf 146} (1966) 858.
  
\bibitem{Lampe:1968zz}
M.~Lampe,
Phys. Rev. \textbf{170} (1968), 306-319
doi:10.1103/PhysRev.170.306
 
 \bibitem{1976ApJ...206..218F}
E.~{Flowers} and N.~{Itoh}, 
  Astrophys. J \textbf{206} (1976) 218-242.
  doi:10.1086/154375.

  
\bibitem{Schmitt:2017efp}
A.~Schmitt and P.~Shternin,
Astrophys. Space Sci. Libr. \textbf{457} (2018), 455-574
doi:10.1007/978-3-319-97616-7\_9
[arXiv:1711.06520 [astro-ph.HE]].




 \bibitem{1984MNRAS.209..511N}
R.~{Nandkumar} and C.~J. {Pethick}, 
 Mon. Not. Roy. Astron. Soc. \textbf{209} (1984) no.4, 511-524
  doi:10.1093/mnras/209.3.511.


\bibitem{Braaten:1989mz}
E.~Braaten and R.~D.~Pisarski,
Nucl. Phys. B \textbf{337} (1990), 569-634
doi:10.1016/0550-3213(90)90508-B

\bibitem{Braaten:1990az}
E.~Braaten and R.~D.~Pisarski,
Nucl. Phys. B \textbf{339} (1990), 310-324
doi:10.1016/0550-3213(90)90351-D

\bibitem{Altherr:1992mf}
T.~Altherr and U.~Kraemmer,
Astropart. Phys. \textbf{1} (1992), 133-158
doi:10.1016/0927-6505(92)90014-Q

\bibitem{LeBellac:1996kr}
M.~Le Bellac and C.~Manuel,
Phys. Rev. D \textbf{55} (1997), 3215-3218
doi:10.1103/PhysRevD.55.3215
[arXiv:hep-ph/9609369 [hep-ph]].

\bibitem{Manuel:2000mk}
C.~Manuel,
Phys. Rev. D \textbf{62} (2000), 076009
doi:10.1103/PhysRevD.62.076009
[arXiv:hep-ph/0005040 [hep-ph]].

\bibitem{Kalikotay:2020snc}
P.~Kalikotay, S.~Ghosh, N.~Chaudhuri, P.~Roy and S.~Sarkar,
Phys. Rev. D \textbf{102} (2020) no.7, 076007
doi:10.1103/PhysRevD.102.076007
[arXiv:2009.10493 [hep-ph]].

\bibitem{Heiselberg:1993cr}
H.~Heiselberg and C.~J.~Pethick,
Phys. Rev. D \textbf{48} (1993), 2916-2928
doi:10.1103/PhysRevD.48.2916

\bibitem{Sarkar:2010bv}
S.~Sarkar and A.~K.~Dutt-Mazumder,
Phys. Rev. D \textbf{82} (2010), 056003
doi:10.1103/PhysRevD.82.056003
[arXiv:1005.1541 [hep-ph]].


\bibitem{Sarkar:2012ww}
S.~Sarkar and A.~K.~Dutt-Mazumder,
Phys. Rev. D \textbf{87} (2013) no.7, 076003
doi:10.1103/PhysRevD.87.076003
[arXiv:1209.5153 [nucl-th]].

\bibitem{Adhya:2013ima}
S.~P.~Adhya, P.~K.~Roy and A.~K.~Dutt-Mazumder,
J. Phys. G \textbf{41} (2014), 025201
doi:10.1088/0954-3899/41/2/025201
[arXiv:1303.6126 [hep-ph]].

\bibitem{Adhya:2012sq}
S.~P.~Adhya, P.~K.~Roy and A.~K.~Dutt-Mazumder,
Phys. Rev. D \textbf{86} (2012), 034012
doi:10.1103/PhysRevD.86.034012
[arXiv:1204.2684 [hep-ph]].



\bibitem{Sharma:2010bx}
    R. Sharma  and S. Reddy
    Phys. Rev. C,    \textbf{83} 2011, 025803
     doi:10.1103/PhysRevC.83.025803
[arXiv:1004.0926 [astro-ph.SR]]

\bibitem{Yakovlev84}
D. G.~Yakovlev, 
Astrophys. and Space Science \textbf{98} (1984), 37-59
doi:10.1007/BF00651950.

 
\bibitem{Akhiezer} A. I. Akhiezer and V. B. Berestetskii, \emph{Quantum Electrodynamics}, \newblock Interscience Publishers, (1965)                                                                               
 \bibitem{Lifshitz} L. P. Pitaevskii and E.M. Lifshitz,  \emph{{Physical Kinetics}}.
\newblock Butterworth-Heinemann, (2012). 
   
  
\bibitem{Chamel:2008ca}
N.~Chamel and P.~Haensel,
Living Rev. Rel. \textbf{11} (2008), 10
doi:10.12942/lrr-2008-10
[arXiv:0812.3955 [astro-ph]].






\bibitem{Rezzolla_book:2013}
L.~{Rezzolla} and O.~{Zanotti}, \emph{Relativistic Hydrodynamics}.
\newblock Oxford University Press, Oxford, UK, 2013.
\newblock 10.1093/acprof:oso/9780198528906.001.0001.

\bibitem{Baym:1971pw}
G.~Baym, C.~Pethick and P.~Sutherland,
Astrophys. J. \textbf{170} (1971), 299-317
doi:10.1086/151216

\bibitem{Nandi:2010fp}
R.~Nandi and D.~Bandyopadhyay,
J. Phys. Conf. Ser. \textbf{312} (2011), 042016
doi:10.1088/1742-6596/312/4/042016
[arXiv:1012.5973 [astro-ph.HE]].

  




 






\end{thebibliography}

\end{document}